\documentclass[twocolumn]{aa}  

\usepackage{graphicx}
\usepackage{txfonts}
\usepackage[colorlinks=true, linkcolor=blue, citecolor=blue, filecolor=blue, urlcolor=blue]{hyperref}

\usepackage{natbib,twoopt}
\bibpunct{(}{)}{;}{a}{}{,}             
\makeatletter
  \newcommandtwoopt{\citeads}[3][][]{\href{http://adsabs.harvard.edu/abs/#3}%
    {\def\hyper@linkstart##1##2{}%
     \let\hyper@linkend\@empty\citealp[#1][#2]{#3}}}
  \newcommandtwoopt{\citepads}[3][][]{\href{http://adsabs.harvard.edu/abs/#3}%
    {\def\hyper@linkstart##1##2{}%
     \let\hyper@linkend\@empty\citep[#1][#2]{#3}}}
  \newcommandtwoopt{\citetads}[3][][]{\href{http://adsabs.harvard.edu/abs/#3}%
    {\def\hyper@linkstart##1##2{}%
     \let\hyper@linkend\@empty\citet[#1][#2]{#3}}}
  \newcommandtwoopt{\citeyearads}[3][][]%
    {\href{http://adsabs.harvard.edu/abs/#3}
    {\def\hyper@linkstart##1##2{}%
     \let\hyper@linkend\@empty\citeyear[#1][#2]{#3}}}
\makeatother

\begin{document} 

   \title{The Host Galaxies of Radio AGN: New Views from  Combining LoTSS and MaNGA Observations}
   \subtitle{}

   \author{Gaoxiang Jin  
          \inst{1}
          \and
          Guinevere Kauffmann
          \inst{1}
          \and
          Philip N. Best
          \inst{2}
          \and
          Shravya Shenoy
          \inst{3} 
          \and
          Katarzyna Małek
          \inst{4} 
          }

   \institute{Max Planck Institute for Astrophysics, Karl-Schwarzschild-Str. 1, D-85741 Garching, Germany\\
              \email{gxjin@mpa-garching.mpg.de}
         \and
             Institute for Astronomy, University of Edinburgh, Royal Observatory, Blackford Hill, Edinburgh EH9 3HJ, UK
        \and
            Centre for Astrophysics Research, University of Hertfordshire, Hatfield, AL10 9AB, UK
        \and
            National Centre for Nuclear Research, Pasteura 7, PL-02093 Warsaw, Poland
        \\
              }
    \date{Received ???; accepted ???}

 
  \abstract
{
The role of radio mode active galactic nuclei (AGN) feedback on galaxy evolution is still under debate. 
In this study, we utilize a combination of radio continuum observations and optical integral field spectroscopic (IFS) data to explore the impact of radio AGN on the evolution of their host galaxies at both global and sub-galactic scales.
We construct a comprehensive radio-IFS sample comprising 5548 galaxies with redshift $z<0.15$ by cross-matching the LOFAR Two-Metre Sky Survey (LoTSS) with the Mapping Nearby Galaxies at APO (MaNGA) survey.
We revisit the tight linear radio continuum - star formation relation and quantify its intrinsic scatter, 
then use the relation to classify 616 radio-excess AGN with excessive radio luminosities over that expected from their star formation rate.
Massive radio AGN host galaxies are predominantly quiescent systems, but the quenching level shows no correlation with the jet luminosity.
The mass assembly histories derived from the stellar population synthesis model fitting agree with the cosmological simulations incorporating radio-mode AGN feedback models.
We observe that radio AGN hosts grow faster than a control sample of
galaxies matched in stellar mass, 
and the quenching age ($\rm \sim 5\,Gyr$) is at larger lookback times 
than the typical radio jet age ($\rm < 1\,Gyr$).
By stacking the spectra in different radial bins and
comparing results for radio AGN hosts and their controls,
we find emission line excess features in the nuclear region of radio AGN hosts.
This excess is more prominent in low-luminosity, low-mass, and compact radio AGN.
The [\ion{N}{II}]/$\rm H\alpha$ ratios of the excessive emission line indicate that radio AGN or related jets are ionizing the surrounding interstellar medium in the vicinity of the nucleus.
Our results support the scenario that the observed present-day radio AGN activity may help their host galaxies maintain quiescence through gas ionization and heating, but it is not responsible for the past quenching of their hosts.
}

   \keywords{galaxies: evolution --
             galaxies: active --
             radio continuum: galaxies
               }

   \maketitle
%

\section{Introduction}\label{sec:intro}
Much effort over the past decade, both in observations and simulations, has been devoted
to the co-evolution of galaxies and their central supermassive black holes (SMBHs).
One of the first pieces of evidence leading to this investigation was the tight correlation between the masses of SMBHs and the stellar velocity dispersion of their host galaxy bulges \citepads[e.g.,][]{1998AJ....115.2285M,2000ApJ...539L...9F}.
Additional indirect evidence supporting the co-evolution scenario is the observation that
the cosmic growth history of SMBHs and the galaxy population is quite similar \citepads[e.g.,][]{2008ApJ...679..118S}.
They both peak at redshift $z\sim2$ and decline rapidly after $z\sim$1 \citepads[see][for a review]{2014ARA&A..52..415M}. These observational results indicate that there are physical processes linking the
growth of the SMBHs and their host galaxies, even though their sizes differ by orders of
magnitude (pc versus kpc).
However, the SMBH-galaxy co-evolution picture is still not fully understood in detail.

Accreting SMBHs are known as active galactic nuclei (AGN), and they are often very luminous with emission across the full spectral energy distribution (SED) \citepads[see][for a review]{2017A&ARv..25....2P}.
These features are believed to arise from energy produced by matter that is accreting onto the
central SMBH. This matter may reside
in an accretion disk and give rise to very strong broadened emission lines from photo-ionized gas
\citepads[BLR, e.g.,][]{1997ASPC..113...80B}, a warm dusty torus contributing to near-to-mid infrared continuum \citepads[e.g.,][]{1987ApJ...320..537B}, outer photoionized gas emitting high excitation narrow lines \citepads[NLR, e.g.,][]{1981PASP...93....5B}, and a corona surrounding the accretion disk that contributes more radiation in X-rays \citepads[e.g.,][]{1989ApJ...346...68S}.
There may also be radio jets created by relativistic charged particles \citepads[e.g.][]{1984RvMP...56..255B}.

The term AGN feedback describes the impact of AGN on the evolution of their host galaxies.
A negative feedback scenario is widely favoured, in which the
energy from accreting material can heat or eject the gas in the host galaxies,
and then prevent star formation globally \citepads[e.g.,][and references
therein]{2012ARA&A..50..455F}.
For most numerical and semi-analytic simulations,
negative AGN feedback is crucial in massive halos to prevent galaxies from
growing too massive \citepads[e.g.,][]{2005Natur.433..604D,2006MNRAS.365...11C}.
On the other hand, the observed AGN-starburst connection in Type II Seyfert galaxies and infrared AGN indicate that positive AGN feedback may also exist \citepads[e.g.,][]{2004ApJ...613..109H}.
The outflows and shocks from AGN may also create dense gas regions and boost star formation \citepads[e.g.,][]{2015A&A...574A..34S,2022Natur.601..329S}.
Alternatively, the correlation between AGN activity and star formation may arise
because star formation and SMBH accretion share a common gas supply \citepads[e.g.,][]{2018MNRAS.478.4238D,2024ApJ...963...99L}.
Recent theoretical models suggest that
the dominant feedback mode transitions from positive to negative as redshift decreases \citepads[e.g.,][]{2022A&ARv..30....6M,2024ApJ...961L..39S}.

In this paper, we focus on the AGN with radio jet features, known as radio AGN.
For this population, kinetic (or jet, radio) mode is considered as the main feedback mechanism.
These AGN mainly output mechanical energy through powerful radio jets, heat the surrounding gas, and produce weak emission lines in the interstellar medium of the host galaxy \citepads[e.g.,][]{2014ARA&A..52..589H}. 
Radio AGN are more likely to be found in massive elliptical galaxies \citepads{2005MNRAS.362...25B}, and in group or cluster environments \citepads{2019A&A...622A..10C} than AGN that emit most of their energy in the
optical or UV.
They usually have low black hole accretion rates and are lack of dusty torus near the black hole \citepads{2014ARA&A..52..589H}.
In galaxy-scale simulations, various AGN jet models have been found, in principle, to be capable of quenching a galaxy and preventing gas from cooling at the centers of dark matter halos \citepads[e.g.,][]{2010MNRAS.409..985D,2016ApJ...829...90Y}.
However, the challenge remains to connect the critical free parameters in these simulations to
observational data.

Radio AGN themselves have diverse observational properties.
They are historically divided into Fanaroff \& Riley \citepads[FR,][]{1974MNRAS.167P..31F} types by radio morphology,
where in general FR\,I, FR\,II, and FR0 \citepads{2019ApJ...871..259G} represent
`edge-darkened', `edge-brightened', and compact jets, respectively.
Recent works suggested that the excitation properties may be more fundamentally connected to the SMBH accretion rate than the morphology \citepads[e.g.,][]{2012MNRAS.421.1569B}.
Radio AGN with strong optical emission lines (e.g., [\ion{O}{III}]$\lambda$5008) are classified as high-excitation radio galaxies (HERGs) with ongoing star formation activity in their hosts. 
The other population without strong emission lines are called low-excitation radio galaxies (LERGs).

The radio flux of normal star-forming galaxies with no AGN is dominated by the synchrotron continuum emitting from relativistic particles moving through magnetic fields.
These particles are mainly accelerated by supernova \citepads{1976A&A....53..295B} and SMBH accretion.
The incidence of supernovae is related to the birth rate of short-lived, high-mass stars ($M>8\rm \, M_{\odot}$),
which is thought to be the origin of the
tight correlation between star formation rate (SFR) and radio luminosity (hereafter radio-SFR relation).
The radio-SFR relation was first observed through the constant ratios of radio and far-infrared luminosity in SFGs \citepads[e.g.,][]{1985A&A...147L...6D,1991ApJ...376...95C},
and has also been confirmed by diverse tracers of SFR
\citepads[e.g.,][Shenoy et al. in prep]{2016MNRAS.461.1898W,2017ApJ...847..136B,2018MNRAS.475.3010G,2021A&A...648A...6S,2022A&A...664A..83H,2023MNRAS.523.1729B,2024MNRAS.531..977D}.
Most radio AGN classification methods are based on the excessive total radio luminosities compared to the star formation properties.
Thus, the excess of radio luminosity compared to the star formation rate is used to derive 
the luminosity of the radio AGN jet.
There are also other approaches to classification using infrared colors and spectral indices \citepads[e.g.,][and reference therein]{2019A&A...622A..17S},
but these methods cannot quantitatively decompose the radio contribution from star formation and AGN jets.

Our current knowledge about the population of radio AGN is based on several all-sky radio continuum surveys at 1.4\,GHz,
such as the NRAO Very Large Array Sky Survey \citepads[NVSS,][]{1998AJ....115.1693C}
and the Faint Images of the Radio Sky at Twenty-Centimeters
\citepads[FIRST,][]{1995ApJ...450..559B}.
In the low-redshift universe ($z<1$), the sensitivity of these surveys probes radio AGN with luminosities
greater than $L\rm _{1.4GHz}\sim 10^{23}\, W\,Hz^{-1}$.

The new generation of radio telescopes and instruments, such as the LOw-Frequency ARray \citepads[LOFAR,][]{2013A&A...556A...2V}, the Australian Square Kilometre Array Pathfinder \citepads[ASKAP,][]{2021PASA...38....9H}, MeerKAT \citepads{2016mks..confE...1J}, etc.,
are now able to discover much fainter radio AGN.
The LOFAR Two Meter Sky Survey \citepads[LoTSS,][]{2017A&A...598A.104S} is one of the deepest ongoing radio continuum large-sky surveys,
which aims to observe the whole northern sky at frequencies 120-168 MHz.
Compared to NVSS and FIRST at 1.4\,GHz,
LoTSS observes at lower frequencies,
which are more sensitive to old, diffuse synchrotron emission
and less affected by the dust continuum thermal emission \citepads[e.g.,][]{2020MNRAS.496.1565P}.
For an extragalactic source with a typical radio spectra index, it is also about one order of magnitude deeper than NVSS and FIRST,
and sensitive enough to detect faint radio sources
which are mainly star-forming galaxies (SFGs) at low redshifts \citepads[][]{2019A&A...622A...2W,2023A&A...678A.151H}.

To study the properties of radio AGN host galaxies,
we need observations at wavelengths where emission from the stellar component and interstellar medium (ISM) become dominant.
The optical integral field unit (IFU) technology is particularly useful to study star formation and 
stellar populations of galaxies.
Spatially resolved galaxy properties, which are not measurable in single-fiber spectroscopic surveys,
are now available from IFU observations.
Within the sky coverage of LoTSS, 
Mapping nearby Galaxies at Apache Point Observatory \citepads[MaNGA,][]{2015ApJ...798....7B}
is the largest  optical IFU survey that has been completed.
MaNGA includes spatially resolved spectroscopic observations of about ten thousand galaxies,
from which we can build a statistically significant sample of radio AGN host galaxies.
The data available for MaNGA galaxies also include classifications of morphology and environment,
and we can thus investigate whether these parameters affect radio AGN properties.

The combination of LoTSS and MaNGA allows us to investigate radio AGN feedback on sub-galactic scales,
particularly for faint radio AGN.
The latest public data release, LoTSS-DR2, and the half MaNGA sample (DR16), were used by
\citetads{2022A&A...665A.144M} to build a sample of 307 radio AGN and to
compare the star formation, ionization maps, and age gradients of radio AGN hosts 
with control galaxies without radio AGN. It was found
that radio AGN are generally quiescent, but are not more quenched than their controls.
The authors concluded that radio AGN help host galaxies maintain quiescence but are not responsible
for switching off star formation completely.
Using a similar sample, \citetads{2023A&A...673A..12Z} studies the stellar kinematics of radio AGN host galaxies.
They found that the low angular momentum is important for galaxies to host luminous radio AGN.

In this paper, we will update the sample using the final MaNGA data release \citepads{2022ApJS..259...35A},
which includes $\sim$6000 galaxies falling within the LoTSS-DR2 sky footprint.
This roughly doubles the size of the radio AGN sample compared to \citetads{2022A&A...665A.144M}.
Firstly, we aim to utilize this currently largest IFU-radio sample to provide an
improved derivation of the radio continuum - SFR relation and its intrinsic scatter.
Because MaNGA provides full-coverage H$\alpha$ maps with dust attenuation information from the Balmer
decrement, we can overcome the limitation of single fiber observations that only
offer nuclear spectroscopic information for low-redshift galaxies. 
This also allows the radio continuum - SFR relation to be better constrained.
Based on this relation, the excess radio luminosity compared to the
prediction from its star formation rate is a physical and reliable criterion to classify radio-excess AGN (hereafter referred to as RDAGN).
This relation is also used to decompose the star formation and AGN contributions to the total radio flux, i.e., to calculate the jet luminosities (or upper limits).

With this RDAGN host galaxy sample in hand, we then aim to study star formation in the RDAGN hosts
and to investigate how the radio jets affect the sub-galactic properties of their hosts.
MaNGA provides maps of the optical spectra (3600\,-\,10300\,\AA) out to 1.5$\times$ effective radius $R_e$.
The emission and absorption lines can be used to analyze the properties of the 
ionized interstellar medium (ISM) and resolved stellar populations within the host galaxy.
We also use results derived from the full spectrum fitting method \citepads[][]{2017MNRAS.472.4297W}
to analyze the star formation histories (SFH) in different regions of the host galaxies.

The structure of this paper is as follows.
In Section~\ref{sec:data}, we introduce the LoTSS-MaNGA parent sample cross-matching and the calculation of physical parameters.
In Section~\ref{sec:agn}, we build the radio continuum - SFR relation
and use it to build RDAGN and  control samples.
In Section~\ref{sec:result}, we examine the global star formation in RDAGN hosts.
Additionally, we analyze the spatially resolved mass assembly history and emission line features of these hosts in different radial bins.
We discuss the main results and the implication of our study as well as the future outlook
in Section~\ref{sec:discuss}.
Section~\ref{sec:conclusion} contains an overall summary of our main results.

Throughout this paper,
we convert all radio luminosity and related relations to rest-frame 144MHz assuming a radio spectral index
$\alpha$=-0.7 \citepads[][]{2002AJ....124..675C}: $L_{\nu_{1}}/L_{\nu_{2}}=(\nu_{1}/\nu_{2})^{-0.7}$.
We adopt a flat cosmology with $H_{0}=70\,\rm km\,s^{-1}\, Mpc^{-1}$, $\Omega_{m}=0.3$, and $\Omega_{\Lambda}=0.7$.
All stellar masses and star-formation rates are based on the Chabrier initial mass function \citepads{2003PASP..115..763C}.
When comparing literature results based on different IMFs,
we follow \citetads{2014ARA&A..52..415M} and set the factor (for stellar masses and star formation rates) as Salpeter\,:\,Kroupa\,:\,Chabrier\,=\,1\,:\,0.67\,:\,0.63.

\section{Data}\label{sec:data}
\subsection{MaNGA}\label{sec:manga}
MaNGA is part of the Sloan Digital Sky Survey IV \citepads[SDSS-IV,][]{2015ApJ...798....7B},
and is the largest completed IFU survey.
The final data products are released in SDSS DR17 \citepads{2022ApJS..259...35A},
which consist of $\sim$10\,000 unique galaxies with good quality IFU coverage out to at least 1.5 times the $r$-band effective radius ($R_{e}$).
MaNGA galaxies are selected from the NASA-Sloan Atlas catalog\footnote{\url{http://www.nsatlas.org/}} \citepads[NSA,][]{2011AJ....142...31B,2017AJ....154...86W}, and
thus have measurements of many physical parameters based on the UV-optical photometric data, such as absolute magnitude, effective radius, position angle, etc. 
We adopt the global stellar mass from the NSA catalog to avoid having to correct for the
fact that the MaNGA IFU sizes vary.
The stellar masses are based on the stellar population synthesis
model fitted from FUV to $z$ band photometry \citepads{2007AJ....133..734B}.
The redshift distribution of the sample is 0-0.15 and peaks at $z\sim0.03$,
while the stellar mass distribution is designed to have a flat distribution between $10^9-10^{11} \rm M_{\odot}$ \citepads{2017AJ....154...86W}.
The IFU cubes have $r$-band FWHM of $\sim$2.5 arcsec and cover 3600\,-\,10300\,\AA \ with a spectral resolution of $R\sim2000$ \citepads{2013AJ....146...32S}.
We note that the MaNGA sample will under-sample at the faint end \citepads[$r$-band absolute magnitude larger than -19,][]{2023ApJ...942..107S}, but this does not affect the main results in this paper since our radio AGN sample is dominated by massive galaxies (see Section~\ref{sec:global}).

All the spatially resolved optical spectra used in this paper are obtained from the MaNGA Data Analysis Pipeline \citepads[{\tt DAP},][]{2019AJ....158..231W,2019AJ....158..160B}.
Emission line, absorption line, and kinematic measurements in this paper 
are taken either directly from {\tt DAP-MAPS} images or fitted from the {\tt DAP-LOGCUBE} spectra.

The global star formation rates are calculated from the attenuation- and AGN-corrected H$\alpha$ luminosity using the relation suggested by \citetads{2012ARA&A..50..531K}:
${\rm SFR\,(M_{\odot}\,yr^{-1})}=L_{\rm H\alpha SF}\, {(\rm erg\,s^{-1})} \times 10^{-41.30}$.
For dust attenuation, we use the Balmer decrement assuming the Case B recombination \citepads{1989agna.book.....O} and the reddening curve from \citetads{2000ApJ...533..682C}: $L_{\rm H\alpha} = L_{\rm H\alpha, obs}\times {\rm [(H\alpha/H\beta)_{obs}/2.86]^{2.6}}$.
The AGN contribution to $L_{\rm H\alpha}$ is estimated by a re-projected optical diagnostic diagram proposed in \citetads{2020MNRAS.499.5749J}, which uses the line ratios [\ion{N}{II}]/H$\alpha$, [\ion{S}{II}]/H$\alpha$, and [\ion{O}{III}]/H$\beta$. The empirical relation can be found in equation 6 of \citetads{2021ApJ...923....6J}.
This step does not affect the star forming galaxy sample used in Section~\ref{sec:l144-sfr}.
For RDAGN sample, the AGN contribution correction changes about -0.2 dex (median value) on the SFRs, while their SFRs are about -2.3 dex below the star formation main sequence,
thus this correction does not affect the discussions in Section~\ref{sec:global} about the global quenching of RDAGN host galaxies.

We build the spatially resolved star formation histories based on the non-parametric, full spectra fitting results from the {\tt Firefly-VAC} \citepads{2017MNRAS.466.4731G,2022MNRAS.513.5988N}. 
The stellar population models \citepads{2020MNRAS.496.2962M} used for fitting are from the MaNGA Stellar library of Milky Way stellar spectra \citepads[{\tt MaStar},][]{2019ApJ...883..175Y}.
These models are also used in the {\tt DAP} and have matched spectral resolution with the galaxy spectra since they are observed by the same instrument.
The age grids of {\tt MaStar} models used in fitting can help us build the mass assembly history of a given spatially resolved region with a precision of at least 1\,Gyr.

\subsection{LoTSS}\label{sec:lotss}
The LOFAR Two Meter Sky Survey (LoTSS) is an ongoing deep wide area radio wavelength imaging survey  \citepads{2017A&A...598A.104S}.
It aims to observe the whole northern sky at frequencies 120-168 MHz.
The expected root-mean-square (RMS) sensitivity and spatial resolution are $\sim$100 $\rm \mu$Jy  (at favorable declinations) and 6 arcsec, respectively.
The latest data release, LoTSS-DR2 \citepads{2022A&A...659A...1S}, covers a 5634 square degrees sky area.
We refer the readers to \citetads{2022A&A...659A...1S} for details about the data reduction.
In summary, LoTSS-DR2 provides well-calibrated 6 arcsec resolution mosaic images.
Additionally, it includes a catalog of approximately 5 million radio sources created by the Gaussian-fitting-based radio source extraction code {\tt PyBDSF} \citepads{2015ascl.soft02007M}.

In this paper, we use the source positions, flux density and spatial size from a value-added catalog with optical host galaxy identification information \citepads[hereafter LoTSS-VAC,][]{2023A&A...678A.151H}.
The host galaxy identification is based on a likelihood-ratio cross-match method described in \citetads{2019A&A...622A...2W},
as well as visual classifications from the citizen science project,  Radio Galaxy Zoo: LOFAR\footnote{\url{http://lofargalaxyzoo.nl/}}.
The optical and infrared information used for this cross-match is based on the DESI Legacy Imaging Surveys\footnote{\url{https://legacysurvey.org/}} \citepads{2019AJ....157..168D} and the unWISE data \citepads{2019ApJS..240...30S}.
About 85\% LoTSS detections have an optical or infrared identification \citepads{2023A&A...678A.151H}.
The total flux density and morphological parameters have been examined,
and where necessary corrected, in the host identification steps.
This correction is necessary for deep radio observations in the MaNGA redshift range,
because many radio galaxies have large sizes or diffuse emissions,
creating large radio-optical offsets which will affect the traditional cross-matching method based on source separations.
Furthermore, the erroneous 
association of a radio galaxy with only one of its multiple components will result in its the radio flux density being significantly underestimated.
For the LoTSS-MaNGA matched sample used in this paper,
the LoTSS-VAC recovers 44 radio galaxies missed in the raw {\tt PyBDSF} catalog and corrects the flux density of 157 galaxies,
which means that the raw catalog would have lost $\sim$8\% accuracy.

\subsection{LoTSS-MaNGA parent sample}\label{sec:crossmatch}
\begin{figure}
\centering
\includegraphics[width=\hsize]{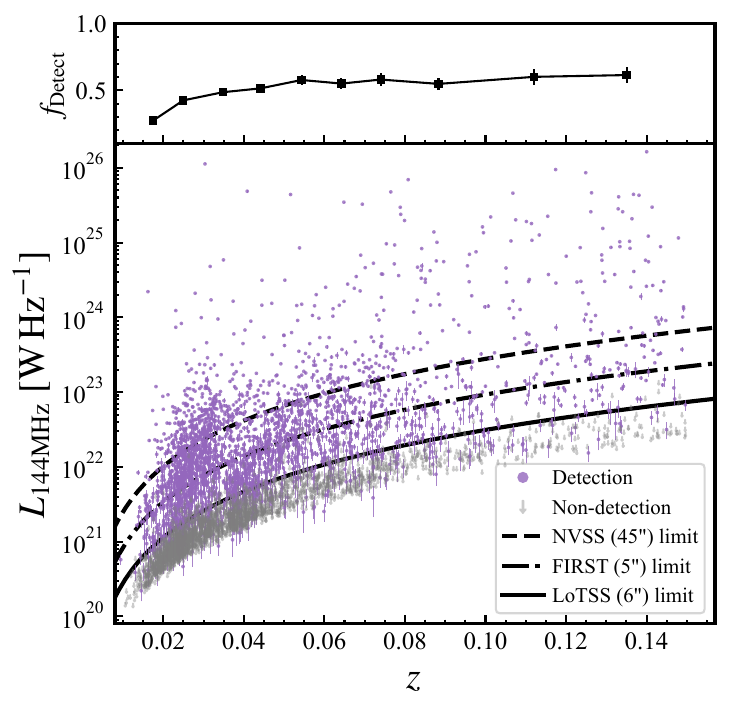}
   \caption{
   Radio luminosity and upper limits of LoTSS-MaNGA sample versus their redshift.
   In the main panel, purple dots show radio luminosities with errors for the LoTSS detections.
   Gray arrows represent the upper limits of the radio luminosity for LoTSS non-detected galaxies.
   Solid line, dash-dotted, and dashed lines represent the 90\% completeness sensitivity limits of LoTSS \citepads{2022A&A...659A...1S}, FIRST \citepads{1995ApJ...450..559B}, and NVSS \citepads{1998AJ....115.1693C}, respectively.
   We note that a few non-detections which are significantly above the LoTSS limit are the sources affected by nearby artifacts. The upper panel shows that the radio detection rate stays approximately constant with redshift, but decreases noticeably at z < 0.03, where the lower mass galaxies start to dominate the MaNGA sample.}
      \label{fig:detectionlimit}
\end{figure}   

Our analysis sample includes all $z>0.01$ MaNGA galaxies (to ensure a large enough IFU field of view) with LoTSS observations.
Galaxies outside the LoTSS sky coverage or in the bad-quality sky areas are excluded based on the mosaic images,
leaving us with 5687 MaNGA galaxies covered by LoTSS DR2.
We match the MaNGA galaxy positions with the galaxy positions in LoTSS-VAC (i.e., the positions for the identified host galaxy, not the radio positions) within a radius of 9 arcsec (1.5$\times$beamsize and smaller than MaNGA IFU)\footnote{
This radius is safe enough since there is only one galaxy or one galaxy merger in a $\sim$20\arcsec \ MaNGA cube, and can also include the sources with large SDSS-unWISE offsets (the FWHM of SDSS images and WISE W1 images are around 2.5\arcsec \ and 6.6\arcsec, respectively). The actual matching offsets are usually far smaller than 9\arcsec, e.g. $\sim$90\% detection have offsets less than 1\arcsec .}.
This process results in 2655 detections (47\%) with radio flux density and size measurements.
We also visually check all the radio and optical images to exclude 109 (2\%) galaxies with severe background source blending or having insufficient MaNGA coverage (IFU coverage is too small).
For 2923 (51\%) non-detections, we measure the median RMS flux
within a 9 arcsec radius aperture around the MaNGA positions in LoTSS images.
We note that there are several non-detections close to luminous radio sources (e.g., MaNGA 8312-3704), and they have much higher RMS levels than sources in the normal fields due to artifacts around the strong sources.
Thus, we consider the 2\% of non-detections with the highest RMS values as outliers and exclude them from our final sample.
We adopt 8 times the RMS flux density in one beam ($F_{\rm RMS}$) as a secure flux density upper limit for each non-detection,
since the catalog source completeness at this cut level reaches more than 95\%
\citepads[][]{2024MNRAS.527.6540H}.

We calculate the rest-frame 144MHz luminosity or upper limit using the formula
$L_{\rm 144MHz} = F_{\rm LoTSS} \times 4\pi D_{L}^{2} \times (1+z)^{(-\alpha-1)}$,
where $D_L$ is the luminosity distance, $z$ is the galaxy redshift from MaNGA observation, and $\alpha$ is assumed radio power-law index (-0.7, see Section~\ref{sec:intro}).
$F_{\rm LoTSS}$ is the total flux density from LoTSS-VAC for detections,
or 8$\times F_{\rm RMS}$ for non-detections.
This rest-frame conversion only corrects $\sim$2\% of the value at MaNGA's peak redshift $z\sim0.03$,
which is smaller than the level of most flux density errors,
thus does not affect our analysis and results.

Fig.~\ref{fig:detectionlimit} shows how the radio luminosity detection
threshold varies with redshift for our sample of MaNGA galaxies with LoTSS-DR2 overlap.
Purple dots represent LoTSS-detected galaxies and grey upper limits are non-detections.
The rest-frame 144MHz luminosity varies between $\rm 10^{20}-10^{26}\,W\,Hz^{-1}$.
LoTSS DR2 reaches 90\% completeness at $\sim$0.8\,mJy \citepads{2022A&A...659A...1S},
resulting in a limit of $L\rm_{144MHz}\sim5\times10^{22}\,W\,Hz^{-1}$ at $z=0.15$.
We note that the exact limiting sensitivity also depends on the sky position, which causes the dispersion for the upper limits in Fig.~\ref{fig:detectionlimit}.
A few non-detections significantly above the LoTSS limit are the sources affected by strong artifacts in the LoTSS images, which are usually caused by nearby luminous sources \citepads{2022A&A...659A...1S}.
We try to minimize the bias from these sources by taking into account both the upper limit and its error when selecting
RDAGN and control samples.
This is discussed in detail in Section~\ref{sec:rdselection} and \ref{sec:control}.

The new LoTSS observations significantly improve the radio detection rate.
As a comparison, 72\% and 41\% of LoTSS detections will fall below the detection limits of NVSS \citepads[dashed line in Fig.~\ref{fig:detectionlimit},][]{1998AJ....115.1693C} and FIRST \citepads[dash-dotted line in Fig.~\ref{fig:detectionlimit},][]{1995ApJ...450..559B}, respectively.
The LoTSS detection rate of MaNGA galaxies (Fig.~\ref{fig:detectionlimit} upper panel) stays approximately constant with redshift, but decreases noticeably at $z<0.03$.
This is due to the fact that the selection of MaNGA parent sample will constrain the apparent sizes of galaxies to not over-fill the IFUs,
thus at lower redshifts the sample will be dominated by physically smaller, and hence lower mass, galaxies which tend to have faint radio emission.

\section{Radio-excess AGN classification}\label{sec:agn}   
\subsection{The $L_{\rm 144MHz}$-SFR relation}\label{sec:l144-sfr}
As mentioned in Section~\ref{sec:intro}, synchrotron continuum luminosity is tightly correlated with the star formation rate in SFGs.
This relation can be used to select radio-excess AGN (or RDAGN for short).
Using LoTSS data, the radio-SFR relation has been investigated based on different SFR tracers and over different spatial scales.
For example, \citetads{2018MNRAS.475.3010G}, \citetads{2021A&A...648A...6S}, and \citetads{2023MNRAS.523.1729B} used SFRs
from SED-fitting for SFG populations out to $z\sim1$,
while \citetads{2022A&A...664A..83H} used SFRs converted from total infrared luminosity for resolved local group SFGs.
Shenoy et al. (in prep) used both SED-fitted SFRs as well as integrated H$\alpha$ derived SFRs.
Up to now, there has been no relation derived from IFU-based H$\alpha$ SFRs, 
despite the higher accuracy of this method.
For consistency, here we derive the $L_{\rm 144MHz}$-SFR relation from the LoTSS-MaNGA sample to help us select RDAGN later.

\begin{figure}
\centering
\includegraphics[width=\hsize]{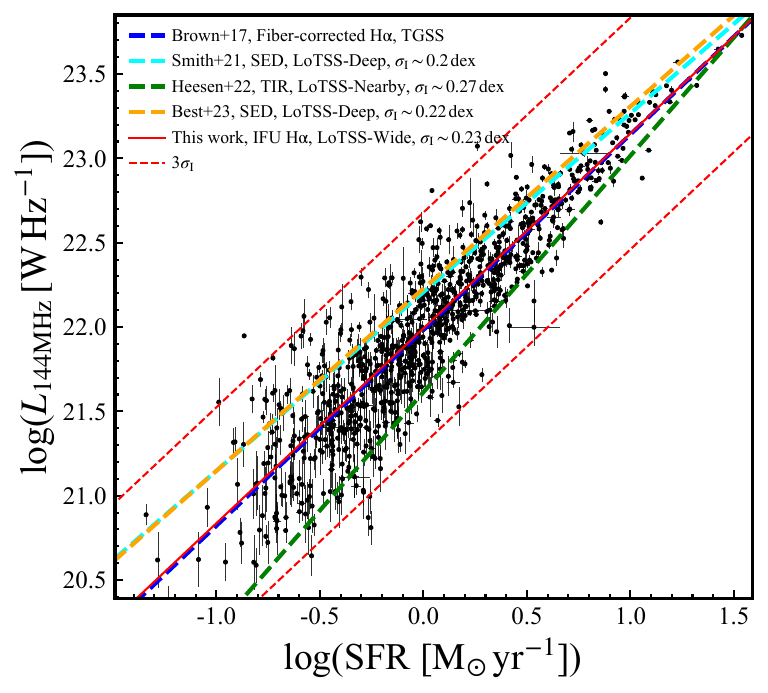}
  \caption{
  The correlation between rest-frame 144MHz luminosity and SFR.
  Black dots with errors are the pure star-forming galaxies used for linear fitting.
  These SFGs are selected based on optical line diagnostic and radio image visual classification (see Appendix~\ref{ap:fit} for details).
  The SFRs are integrated from MaNGA H$\alpha$ maps and corrected for dust attenuation as mentioned in Section~\ref{sec:manga}.
  The red solid line is the MCMC linear fitting result.
  Intrinsic errors, $\pm 3 \sigma_{\rm I}$, are shown in red dashed lines.
  Blue, cyan, green, and orange dashed lines are the  linear fitting results in \citetads{2017ApJ...847..136B}, \citetads{2021A&A...648A...6S}, \citetads{2022A&A...664A..83H}, and \citetads{2023MNRAS.523.1729B}, respectively.
  Our result is in good consistency with most literature results.
   }
     \label{fig:l144-sfr}
\end{figure}

The details of how we derive this relation can be found in Appendix~\ref{ap:fit}.
In brief, we select a pure star-forming galaxy sample without AGN contamination
of both the H$\alpha$ and radio continuum luminosities.
Then the $L_{\rm 144MHz}$-SFR relation is derived by performing a linear fit to the rest-frame 144MHz luminosities and SFRs, inspired by the efforts in \citetads{2018MNRAS.475.3010G}.
The intrinsic errors $\sigma_{\rm I}$ of this relation are determined using the {\tt emcee} \citepads{2013PASP..125..306F} Markov chain Monte Carlo (MCMC) method.

Our derived radio-SFR relation can be expressed as Equation~\ref{eq:l144-sfr}.
\begin{equation}\label{eq:l144-sfr}
{\rm log}(\frac{L_{\rm 144MHz}}{\rm W\,Hz^{-1}})=1.16\times {\rm log}(\frac{\rm SFR}{\rm M_{\odot}\,yr^{-1}}) + 21.99 \pm 0.23
\end{equation}
Fig.~\ref{fig:l144-sfr} shows that our results generally agree with previous low redshift relations based on different samples.
For comparison,
\citetads{2017ApJ...847..136B} (blue line) used shallow TGSS sample with SFRs from single fiber H$\alpha$;
\citetads{2021A&A...648A...6S} (cyan line) and \citetads{2023MNRAS.523.1729B} (orange line) used a sample with deep radio observations and SFRs from multi-wavelength SED fitting;
and \citetads{2022A&A...664A..83H} (green line) used a sample of local galaxies with SFRs measured from total infrared luminosities.
Our results suggest an intrinsic scatter of $\sigma_{\rm I} \sim0.23$ dex in this relation,
consistent with other LoTSS samples \citepads[e.g.,][]{2021A&A...648A...6S,2023MNRAS.523.6082C}.
The slope is larger than one, which supports the prediction of calorimetry models \citepads[e.g.,][]{1990MNRAS.245..101C}.

This relation was found to also depend on the stellar mass \citepads[e.g.,][Shenoy et al. in prep]{2021A&A...648A...6S,2022A&A...664A..83H},
but the mass dependence is weak and only significant in low-mass systems \citepads{2024MNRAS.531..977D} and at higher redshift \citepads{2021A&A...647A.123D}, 
thus we do not include the mass parameter in our fitting.
Our fitting procedure also does not include the radio non-detections, which may result in a slightly more conservative cut for radio AGN classification.
Though our result is reliable enough to classify bona fide radio AGN in massive galaxies (e.g. $M_* > 10^{9.5} \rm M_{\odot}$), readers should be careful and test the mass dependence when applying it in lower mass samples.

\subsection{Radio-excess AGN and their jet luminosities}\label{sec:rdselection} 

We classify a MaNGA galaxy as a radio-excess AGN host if its $L\rm _{144MHz}$ is at least $3\sigma$ level higher than the $L\rm _{144MHz}$ predicted by its SFR and Equation~\ref{eq:l144-sfr}.
Considering the uncertainty of these measurements, the classification can be expressed as follows:
${\rm log}(L_{\rm 144MHz} - 3\sigma _{L}) > 1.16 \times \rm log(SFR + 3\sigma _{SFR}) + 21.99 +  3\sigma _{I}$.
This method results in a new sample of 616 (11\%) radio AGN hosts among 5548 LoTSS-MaNGA galaxies.
This way the AGN sample should all be secure AGN, but it may be incomplete at lower radio luminosities.
Other detections and non-detections are considered as galaxies without secure evidence of radio excess (hereafter referred to as `non-RDAGN' for convenience)."

We also calculate the radio jet luminosity $L_{\rm AGN}$
by subtracting the star formation contribution from the total radio luminosity: $L_{\rm AGN}=L_{\rm 144MHz}^{\rm observed} - L_{\rm 144MHz}^{\rm SFR}$,
where $L_{\rm 144MHz}^{\rm SFR}$ is the 144MHz luminosity predicted by Equation~\ref{eq:l144-sfr}.
The error on the $L_{\rm AGN}$, $\sigma _{L_{\rm AGN}}$ estimate
is derived from the observed uncertainty in $L_{\rm 144MHz}^{\rm observed}$ and the uncertainty in $L_{\rm 144MHz}^{\rm SFR}$,
and thus includes both $\sigma_{\rm SFR}$ and $\sigma _{\rm I}$.

For non-RDAGNs, we need an estimate of the upper limit of $L_{\rm AGN}$ to build a reliable control sample. 
The SMBHs in non-RDAGN may still have weak radio activities hidden in the host star formation\footnote{For example, we can consider a galaxy which has $L_{\rm 144MHz}=10^{23}\rm \,W\,Hz^{-1}$ with a $L_{\rm AGN}=10^{22}\rm \,W\,Hz^{-1}$ radio AGN inside it. It will be classified as non-RDAGN because the radio AGN is hidden in the star formation. This way it should not be selected as a control galaxy of a RDAGN with $L_{\rm AGN}=10^{22}\rm \,W\,Hz^{-1}$, and the $L_{\rm AGN}$ upper limit can help us to quantify this effect.}.
For non-RDAGN with radio detections, we consider that the radio continuum is dominated by star formation rather than AGN,
thus we assume a maximum upper limit of $L_{\rm AGN} + 3\sigma _{L_{\rm AGN}}$.
For non-RDAGN not detected by LoTSS,
we adopt the upper limit of radio luminosity (defined in Section~\ref{sec:crossmatch}) also as the upper limit of $L_{\rm AGN}$.
By these definitions, the non-RDAGN's $L_{\rm AGN}$ upper limit is the maximum possible luminosity of the `hidden AGN'.
A non-RDAGN with the $L_{\rm AGN}$ upper limit much lower than the $L_{\rm AGN}$ of a RDAGN can thus be confidently selected as a control galaxy.
We note that the high SFR star-forming galaxies (SFGs) with radio detection will then have higher $L_{\rm AGN}$ upper limits, because they can have more radio luminosity dominated by star formation than some low SFR RDAGN.
The consequences of this selection effect are discussed in Section~\ref{sec:bias}.

There are two existing MaNGA radio AGN samples in the literature \citepads{2020ApJ...901..159C,2022A&A...665A.144M}.
The key improvements that we have implemented in this work are,
1) classifying AGN based on the self-consistent MaNGA-derived $L_{\rm 144MHz}$-SFR relation,
2) including new measurements of $L_{\rm AGN}$ as well as upper limits in the case of non-detections.
Our sample is therefore better suited to select a non-radio AGN control sample with quantified $L_{\rm AGN}$ (or upper limits)
and is more accurate at distinguishing some high SFR, radio-luminous SFGs from RDAGN.
Compared to \citetads{2020ApJ...901..159C},
where the sample is matched to the NVSS/FIRST selected radio AGN catalog \citepads{2012MNRAS.421.1569B},
the new deep LoTSS observations that include many more star-forming galaxies
allow us to use the more physical $L_{\rm 144MHz}$-SFR selection rather than the $L_{\rm 1.4GHz}$-D$_{\rm n}$4000 classification proposed in \citetads{2005MNRAS.362....9B}.
In \citetads{2022A&A...665A.144M}, 
the radio AGN sample is selected from LoTSS DR2,
but the radio classification uses an empirical relation based on a non-IFU sample \citepads{2019A&A...622A..17S}.

\begin{figure}
\centering
\includegraphics[width=\hsize]{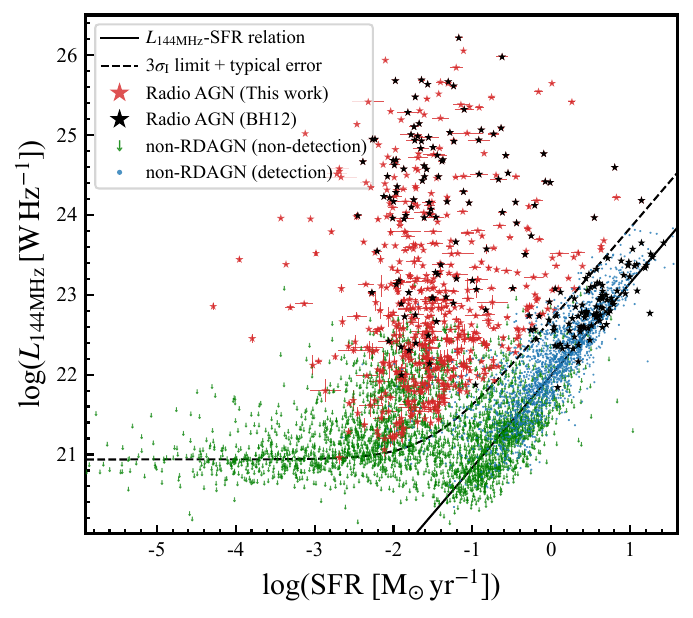}
\caption{
Radio AGN classification for the whole LoTSS-MaNGA sample. Black solid and dashed lines are the radio-SFR relation (Equation~\ref{eq:l144-sfr}) and the $3\sigma_{\rm I}$ limit plus the typical $L_{\rm 144MHz}$ error}, respectively. Detections with observed $L_{\rm 144MHz}$ larger than the limit are classified as radio-excess AGN (red stars). Other detections (blue) and non-detections (green) are considered as radio normal galaxies (non-RDAGN).
We plot the matched radio AGN classified in \citetads{2012MNRAS.421.1569B} and \citetads{2020ApJ...901..159C} as a comparison (black stars).
Our sample includes more low luminosity radio AGN and can distinguish some high SFR SFGs, and 77\% of radio AGN in our sample are newly classified.
 \label{fig:AGNelect}
\end{figure}

Fig.~\ref{fig:AGNelect} illustrates our radio AGN classification for the entire LoTSS-MaNGA sample. Detections above the $3\sigma_{\rm I}$ limit are classified as radio-excess AGN (red),
while other detections (blue) and non-detections (green) are classified as non-RDAGN with an upper limit for jet luminosity.
We also show the radio AGN classified in \citetads{2012MNRAS.421.1569B} using NVSS and FIRST data, i.e., the \citetads{2020ApJ...901..159C} sample
(black stars),
demonstrating not only that our sample probes radio AGN to significantly lower luminosities, but also that some high SFR SFGs may be misclassified as AGN using their methods.
77\% of radio AGN in our sample are newly classified.

\subsection{Control sample selection}\label{sec:control} 
\begin{figure*}
\centering
\includegraphics[width=\hsize]{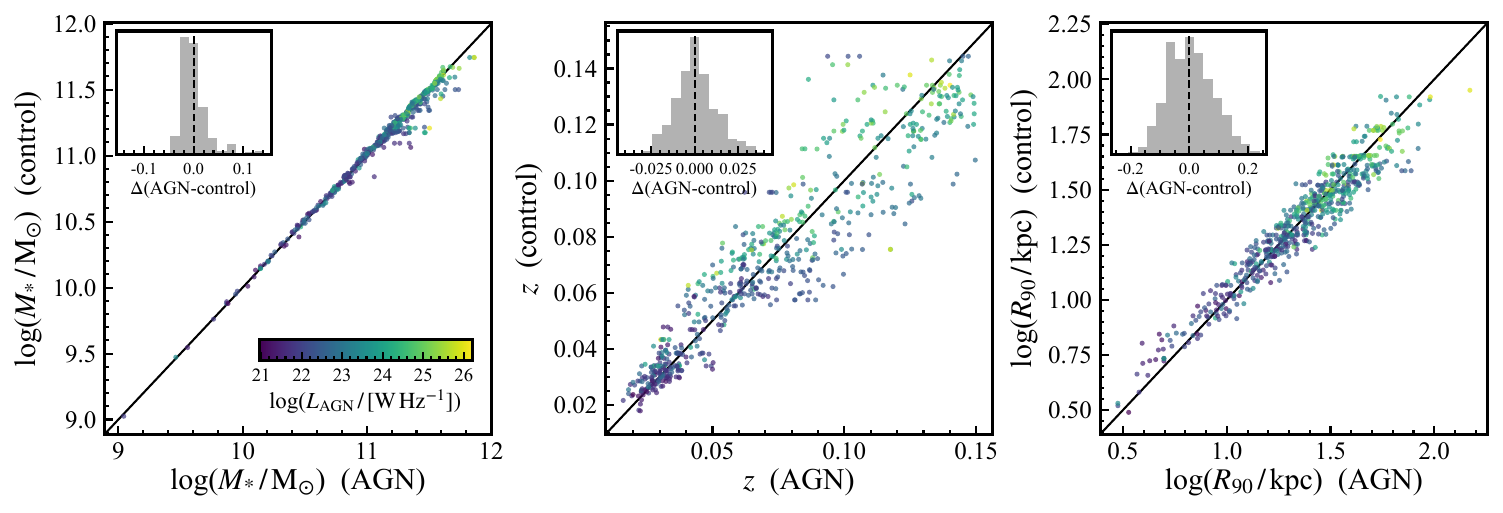}
\caption{Left panel: The stellar mass of each radio AGN host, $M_*^{\rm AGN}$, compared to the mean stellar mass of its control sample, $M_*^{\rm control}$. The black dashed line is the 1:1 relation. $M_*^{\rm control}$ matches with $M_*^{\rm AGN}$ for most galaxies. 
A few galaxies with only one control galaxy have slightly larger offsets, but our scientific results are not affected.
The middle and right panels show similar comparisons for the redshift and 90 percent light radius. Both parameters are also well controlled.
We note that the matched $M_*$ and redshift originate directly from the selection criteria in Section~\ref{sec:control}, while the matched  $R_{90}$ is the result of the matched angular distance and physical effective radius.}
\label{fig:control}
\end{figure*} 

It is well-established that galaxy age and star formation rate (SFR) are strongly correlated with stellar mass \citepads[e.g.,][]{2005MNRAS.362...41G,2015ApJ...808L..49G}.
To evaluate possible different star formation histories of radio AGN host galaxies, 
it is essential to construct mass-controlled samples for comparison. 
For each radio AGN host, we create a control sample consisting of non-RDAGN (no confirmed radio excess) that meet the following criteria:
\begin{itemize}
\item Jet luminosity:  upper-limit$(L_{\rm AGN}^{\rm control}) <  0.2 \times L_{\rm AGN}$,
\item Stellar mass: |log($M_{*}^{\rm AGN}/M_{*}^{\rm control}$)| $<$ 0.3,
\item Physical effective radius: |log($R_{e}^{\rm AGN}/R_{e}^{\rm control}$)| $<$ 0.2,
\item Angular diameter distance: |log($D_{A}^{\rm AGN}/D_{A}^{\rm control}$)| $<$ 0.2.
\end{itemize}

The first criterion ensures that even if the control galaxy contains an  unclassified radio AGN , its maximum jet luminosity  is still marginal compared to that of the corresponding RDAGN (at most 20\%).
Other criteria are set to select as many control sample as possible while considering the uncertainty level of the measurements. For instance, the stellar mass calculation results from different methods are found to have a scatter of 0.2-0.3\, dex \citepads[e.g.][]{2018A&A...620A..50M,2022MNRAS.513.5988N,2022ApJS..262...36S},
the scatters among the $R_{e}$ fitted from different profiles (e.g. Petrosian and Sérsic) are about 0.1-0.2\, dex.

Since radio AGN tend to be more massive than non-RDAGN,
there are typically more less-massive galaxies available in the control sample for a given AGN. 
To maintain a roughly equal median stellar mass between the AGN and control samples,
for each AGN we select an equal number of more massive galaxies and less massive
galaxies that meet the above criteria.
Out of the 616 radio AGN, only three lack any control galaxies that meet these criteria and are discarded.

The left panel of Fig.~\ref{fig:control} illustrates the comparison of the stellar masses of each radio AGN ($M_*^{\rm AGN}$) and the mean stellar masses of their control samples ($M_*^{\rm control}$). 
469 (77\%) radio AGN have at least four control galaxies.
Most radio AGN hosts exhibit matched stellar mass with their control samples.
We also plot a similar comparison for the redshift and 90\% light radius in Fig.~\ref{fig:control},
which shows that the RDAGN share consistent properties with their control sample.
In the left panel of Fig.~\ref{fig:control}, the few dots with slightly larger $M_*^{\rm AGN}-M_*^{\rm control}$ offsets are the radio AGN hosts with only one control galaxy.
These galaxies constitute a small fraction (14\%),
and their offsets are less than 0.3\,dex,
thus not significantly impacting the scientific results of this study.

We do not constrain the control sample to have similar SFR with the RDAGN hosts,
because our focus is on the star formation history dating back to the the quenching of radio AGN host galaxies. Therefore, SFR is considered a free parameter.
Nevertheless, the median difference in SFR between radio AGN hosts and their control sample is only $\sim$0.23\,dex,
which is marginal compared to the standard deviation of the radio AGN SFRs ($\sim$1\,dex).

\section{Results}\label{sec:result}

\subsection{Radio AGN prefer massive quiescent hosts}\label{sec:global}

\begin{figure}
\centering
\includegraphics[width=\hsize]{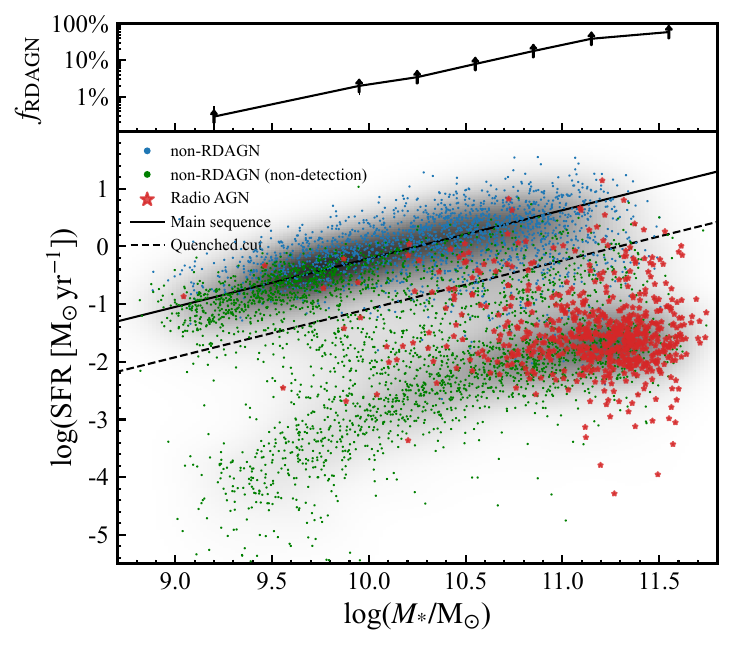}
   \caption{
   The main panel is the SFR versus stellar mass diagram for the LoTSS-MaNGA sample.
   Background gray-scale density map shows the bimodal distribution of star formation main sequence and the quenched population.
   Radio AGN are plotted as red stars while galaxies without confirmed radio excess (non-RDAGN) are plotted as blue (detection) and green (non-detection) dots.
   The black solid line is the star formation main sequence (SFMS).
   Galaxies below the 3$\sigma_{\rm SFMS}$ lower limit of SFMS (black dashed line) are considered as `quenched'.
   Radio AGN host galaxies are dominated by massive and quenched populations.
   The upper panel shows the fraction of radio AGN in all quenched galaxies. The fraction increases with stellar mass to $>50$\% at high mass end ($\rm 10^{11.5}\,M_{\odot}$).}
      \label{fig:sfms}
\end{figure} 

\begin{figure}
\centering
\includegraphics[width=\hsize]{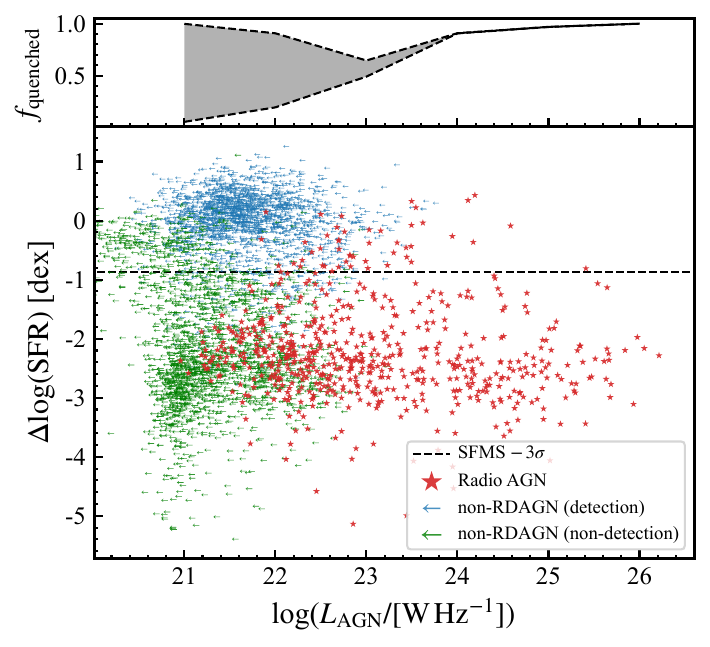}
   \caption{
   The main panel presents the quenching level ($\rm \Delta SFR$ compared to the main sequence SFR at the same stellar mass) of MaNGA galaxies as a function of their radio AGN luminosities (or the upper limits).
   The $L\rm _{AGN}$ upper limits of non-detections and are plotted as green arrows.
   The upper limits of non-RDAGN detections are calculated based on the star formation contribution to the total radio luminosity and are shown in blue arrows.
   Below the black line (3$\sigma$ lower than the SFMS) are quenched galaxies.
   The lack of galaxies in the upper right panel indicates that almost all the hosts of luminous radio AGN are quenched.
   At $L\rm _{AGN}>10^{23}\, W\,Hz^{-1}$, the fraction of radio AGN hosts that are quenched galaxies, $f_{\rm quenched}$,
   increases with $L\rm _{AGN}$ from $\sim$50\% to 100\%,
   as shown in the upper panel.
   The uncertainty shown as the gray region is due to the $L\rm _{AGN}$ limits.
   $\rm \Delta log(SFR)$ shows no correlation with $L\rm _{AGN}$ (Pearson correlation coefficient $\rho \sim -0.11$).
   }
      \label{fig:dsfms}
\end{figure} 

Fig.~\ref{fig:sfms} illustrates the global star formation rate of our RDAGN sample.
The majority of RDAGN hosts (represented by red stars) are situated below the star-forming galaxy
main sequence (SFMS, depicted by the black solid line).
Galaxies below the 3$\sigma_{\rm SFMS}$ lower limit of the SFMS are defined as `quenched',
as they have transitioned from the SFMS to the quiescent population.
Notably, the RDAGN hosts are predominantly composed of quenched galaxies (567 of 616, 92\%).
Furthermore, we investigated the occurrence of RDAGN in the quenched galaxy population.
As shown in the upper panel of Fig.~\ref{fig:sfms},
the fraction of RDAGN among all quenched galaxies,
$f_{\rm RDAGN}$, exhibits an increasing trend with stellar mass,
reaching at least 50\% at $M_{*}\sim 10^{11.5}\rm M_{\odot}$,
which indicates increased recurrence of RDAGN in massive galaxies.
We note here that this fraction $f_{\rm RDAGN}$ can only be considered as a lower limit at the LoTSS sensitivity,
because LoTSS is still not deep enough to confirm if some weak radio excess exists in these non-detected galaxies.
Deeper radio continuum surveys in the future will probe fainter radio AGN.

These results are consistent with our current knowledge about RDAGN in the local universe.
Previous studies have indicated a preference for RDAGN to be hosted in massive, old, and elliptical galaxies \citepads{2005MNRAS.362...25B}.
Recent LOFAR data also suggest that locally, all the $M_{*} > 10^{11}\rm M_{\odot}$ massive galaxies have a RDAGN inside \citepads{2019A&A...622A..17S}.
The fraction observed in our sample is considerably smaller (41\%),
but is still reasonable considering that we use very different RDAGN classifications and the fraction definition compared to theirs.
These results all suggested that the radio AGN hosts currently have old stellar populations and have an early and fast assembly history, which we will investigate in detail in Section~\ref{sec:sfh}.

In Fig.~\ref{fig:dsfms}, we explore if there is a relationship between global quenching and the power of radio jets.
We quantify the degree of quenching by measuring how far galaxies lie below the SFMS: $\rm \Delta log(SFR) = log(SFR_{observed}) - log( SFR_{SFMS})$, i.e. smaller (more negative) $\rm \Delta log(SFR)$ indicates a more quenched galaxy.
Galaxies below the 3$\sigma$ threshold of the SFMS (the black dashed line) are classified as quenched.
The upper panel of Fig.~\ref{fig:dsfms} displays the fraction of quenched galaxies among all RDAGN hosts ($f_{\rm quenched}$) across different $L_{\rm AGN}$ bins.

Due to the nature of the radio luminosity function of SFGs and radio AGN \citepads[e.g.][]{2019A&A...622A..17S}, the completeness of the RDAGN sample will decrease at low $L_{\rm AGN}$, where faint RDAGN in luminous SFGs will become indistinguishable.
At $L_{\rm AGN} > 10^{24}\, \rm W\,Hz^{-1}$, all the radio sources are confidently classified as RDAGN (see Figure \ref{fig:AGNelect}), and
$\sim$95\% (140 out of 147) of them are below the $-3\sigma$ limit of the SFMS,
indicating that nearly all hosts of luminous RDAGN are quenched.
For less luminous RDAGN,
the range of $f_{\rm quenched}$ (the shadow region in the upper panel)
is determined by considering two extreme assumptions.
First, as the lower limit that all $L_{\rm AGN}$ upper limits indicate that their hosts do not host RDAGN.
Second, as the upper limit that all hosts host radio jets with $L_{\rm AGN}$ below the upper limits.
Even under the first assumption, the fraction remains at least $\sim$50\% down to AGN luminosities of $\sim 10^{23}\, \rm W\,Hz^{-1}$.
These fractions support the hypothesis that there are connections between the existence of RDAGN and the quiescence of their hosts.
However, the Pearson coefficient between $\rm \Delta log(SFR)$ and $L_{\rm AGN}$ is approximately -0.1,
indicating that the quenching level is irrelevant to the radio AGN power.

We note that some radio-detected galaxies have $L_{\rm AGN}$ upper limits in Fig.~\ref{fig:dsfms} (blue arrows).
This is because star formation dominates the total radio luminosity in those galaxies, making it hard to classify weak radio AGN based on radio luminosity.
For these galaxies, the LoTSS sensitivity is no longer the limitation to confirm if there is radio AGN activity.
Higher resolution star formation and radio continuum maps would be needed to decompose the AGN jets from star formation at sub-galactic scales.

\begin{figure*}
\centering
\includegraphics[width=\hsize]{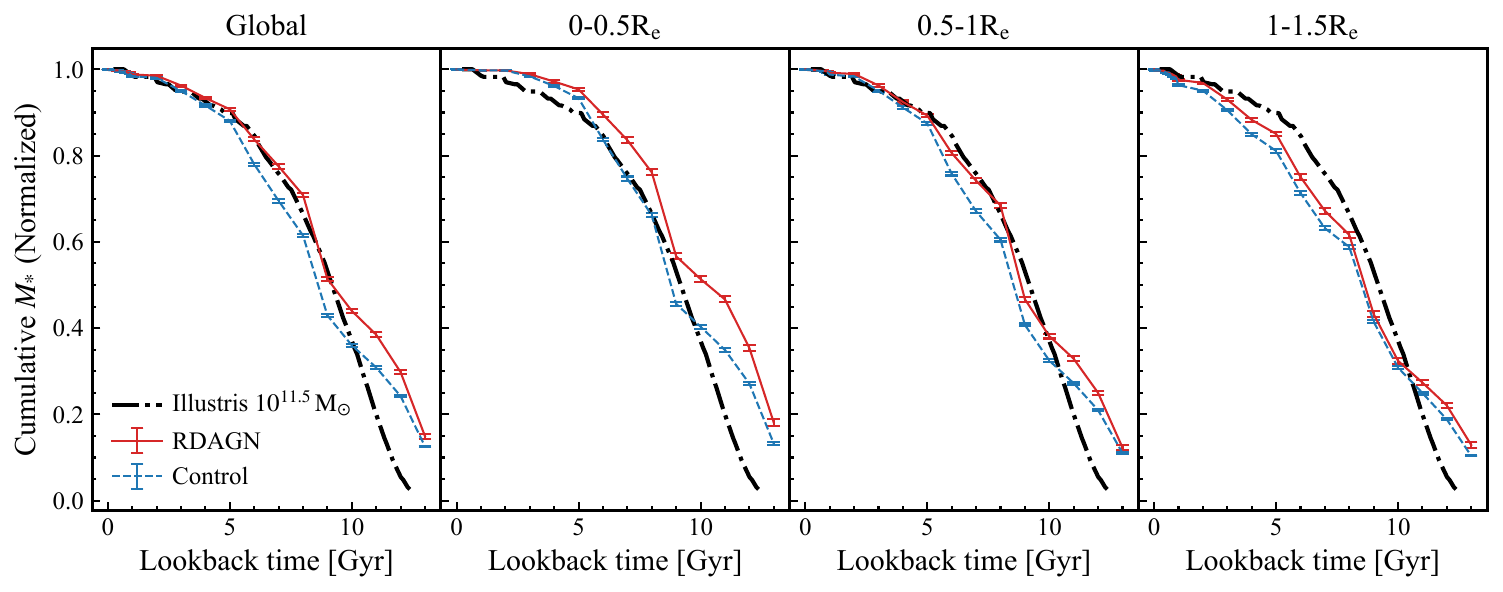}
  \caption{
  Mass assembly across the cosmic lookback time for different regions of RDAGN hosts and their controls.
  Solid red lines are the median SFHs of radio AGN hosts and the blue dashed lines are for their control samples' median.
  We also plot the average mass assembly history of galaxies with halo masses of $\rm 10^{13}\,M_{\odot}$ from the Illustris simulation \citepads{2016MNRAS.458.2371R}. Although the simulation relation is for the total stellar mass, it
  is kept fixed in all the panels for reference.
  Panels from left to right are the results for the regions of the whole galaxy, 0-0.5\,$R_{e}$, 0.5-1\,$R_{e}$, and 1-1.5\,$R_{e}$, respectively.
  RDAGN hosts formed faster than their controls at all radii, especially in the central region.
  Our spectra-fitted  total mass assembly history is reasonably  matched by the cosmological simulation which incorporates the radio AGN feedback model.
  }
     \label{fig:sfh}
\end{figure*} 

In summary, quiescent galaxies dominate the RDAGN hosts, 
with $f_{\rm quenched}$ increasing with $L_{\rm AGN}$ while $\rm \Delta log(SFR)$ shows no correlation with $L_{\rm AGN}$.
`Radio-mode' AGN feedback is believed to be necessary to regulate the cooling flows
in dense cluster environment and help massive galaxies maintain quiescent \citepads[e.g.,][]{2012ARA&A..50..455F}.
Radio-mode AGN feedback models have been implemented in cosmological hydrodynamical simulations,
where the existence of jet mode feedback is crucial for preventing galaxies from becoming too massive
\citepads{2014MNRAS.438.1985T}.
Because the jets dump most of their energy at large distances from the galaxy and because the
radio jet power varies on very short timescales compared to the cooling time of the gas
in dark matter halos, the luminosity of the jet may not
correlate very strongly with the star formation rate in the galaxy.
In addition, several observational studies have found weak correlations between
the radio jet luminosity and both the kinetic power  of the jet and the SMBH accretion rate \citepads[e.g.,][]{2012ARA&A..50..455F,2013MNRAS.432..530R},
suggesting that the energy transfer from radio jets to surrounding gas is inefficient.
These factors may explain why the radio luminosity of our RDAGN sample is irrelevant to the quenching level of the host galaxy.

\subsection{Radio AGN host galaxies quenched earlier}\label{sec:sfh}

In this subsection, we investigate the mass assembly history of the RDAGN host galaxies.
It has been widely argued in simulations that
negative AGN feedback effects are necessary to halt the growth of massive galaxies \citepads[e.g.,][]{2000MNRAS.311..576K,2005MNRAS.361..776S,2015MNRAS.446..521S,2019MNRAS.490.3234N}.
If negative feedback is universally present in real galaxies, we should observe AGN-related suppression of star formation in the star formation histories (SFHs) of galaxies.

As mentioned in Section~\ref{sec:manga}, we employ the full-spectrum fitting method to obtain
radially-resolved non-parametric SFHs.
In Section~\ref{sec:global} we demonstrated that RDAGN hosts are mostly massive quiescent galaxies,
with predominantly old stellar populations that are insensitive to recent star formation.
To better illustrate the SFHs,
we derive the mass accumulation history instead of the SFR history
by quantifying the fraction of a galaxy’s stellar mass that was formed at different lookback times.
We construct the mass assembly histories for each RDAGN host and their control galaxies in three radial bins: 0-0.5\,$R_{e}$ to represent the unresolved central regions, 0.5-1\,$R_{e}$ for the middle regions, and 1-1.5\,$R_{e}$ for the outskirts.
The mass assembly history of individual galaxies can be somewhat uncertain,
so we combine them to compare the average  difference between radio galaxies and the control sample.
The mass assembly histories are normalized at the `observed time' of $t_{\rm lookback}=0$,
corresponding to a median redshift of $z\sim0.07$ (0.9\,Gyr before today).
For each RDAGN host, the controlled history is the median of its control sample's SFHs.
Then the median of all controlled histories are considered as the final results, and uncertainties are estimated using the error of the median (standard deviation divided by the square root of the sample size).

Fig.~\ref{fig:sfh} shows the comparison of median mass assembly histories between RDAGN hosts and their controls.
For both radio AGN hosts (red solid lines) and their controls (blue dashed lines),
galaxies form the central region (0-0.5\,$R_{e}$) earlier than the middle region (0.5-1\,$R_{e}$) and outskirts (1-1.5\,$R_{e}$), following an `inside-out' formation scenario.
At all radii, RDAGN built their mass faster and quenched earlier,
with the early quenching more pronounced in the central region.
Note that this difference does not necessarily imply
a direct connection between RDAGN and the evolution of their hosts, because of the 
very different timescales of radio jet evolution compared to the buildup of the stellar mass.
It is possible that different conditions prevail during the bulk of the stellar mass build-up in radio AGN host galaxies and the control galaxies.
Faster growth could be a consequence of radio AGN being located in denser environments \citepads[e.g.,][]{2010ApJ...721..193P}.
This scenario is also favored by the more significant impact in the central region; in the standard $\Lambda$CDM
cosmological paradigm where structure forms hierarchically, the formation time of the inner region
of the galaxy is boosted to earlier epochs in dense environments.
RDAGN activity may only be triggered after their hosts exceed a certain mass threshold and may 
only play an important role in maintaining the quiescence
at a late stage of the host galaxies' evolution.
Such a scenario is in keeping with the implementation
of radio mode feedback in semi-analytical models of galaxy formation \citepads[e.g.,][]{2006MNRAS.365...11C}.

Our SFHs derived from stellar population synthesis fitting are consistent with some simulations including radio AGN feedback models.
In Fig.~\ref{fig:sfh}, we plot the mass assembly history of $\rm \sim 10^{11.5}\,M_{\odot}$ galaxies in a $\rm 10^{13}\,M_{\odot}$ halo from the Illustris cosmology simulation \citepads{2016MNRAS.458.2371R}.
This simulation includes a `radio-mode' AGN feedback model that is found to be important in suppressing star formation in massive galaxies.
These authors also found that strong and weak radio jets create similar impacts on the result \citepads{2014MNRAS.438.1985T},
which is consistent with the lack of correlation between radio luminosity and the quenching level,
as discussed in Section~\ref{sec:global}.
The mass assembly of the simulated galaxies (black line in Fig.~\ref{fig:sfh}) generally agrees 
with our observational results.
The discrepancy between simulated and spectra-derived SFHs at $t\rm _{lookback}>10\, Gyr$ is due to that the stellar population synthesis is not sensitive to the oldest stars, i.e. the SFH at early universe.

\begin{figure}
\centering
\includegraphics[width=\hsize]{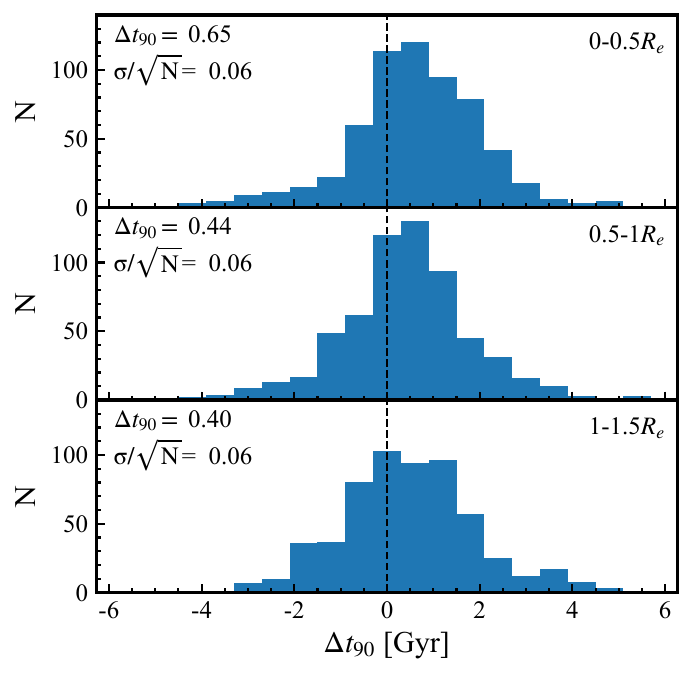}
  \caption{Distributions of the difference between the $t_{90}$ from RDAGN and their controls. 
  Three panels from top to bottom are the results for the inner, middle, and outer regions of galaxies.
  The median difference and its error (standard deviation divided by the square root of the sample size) are noted in the top left of each panel.
  Larger $t_{90}$s indicate that RDAGN hosts globally become quenched earlier.
  }
     \label{fig:t90}
\end{figure}

With the mass accumulation curves,
we calculate the lookback time $t_{\rm X}$ when the galaxy accumulates X percent of the current mass.
These lookback times are often used to quantify how quickly galaxies quench or how early galaxies become quenched \citepads[e.g.,][]{2016ApJ...824...45P}.
Since our sample predominantly consists of quiescent galaxies with no significant recent star formation, we consider $t_{90}$ as the lookback time when galaxies cease rapid growth.
The difference can be determined through the values from RDAGN hosts and their control sample's median values, as $\Delta t_{90}$ (positive means that the $t_{90}$ of RDAGN hosts is larger, i.e., these galaxies formed earlier).
We calculate the $\Delta t_{90}$ for each RDAGN host compared to its controls,
and plot the distributions in Fig.~\ref{fig:t90}.
Panels in Fig.~\ref{fig:t90} from top to bottom are the results in radial bins 0-0.5\,$R_{e}$, 0.5-1\,$R_{e}$, and 1-1.5\,$R_{e}$, respectively.
The median values and errors on the median (the standard deviation divided by the square root of the sample size) are listed in the top left of each panel to indicate the significance of the differences.
In Fig.~\ref{fig:t90}, the distribution of $\Delta t_{90}$ 
is shifted towards positive values in all three radial bins, especially in the central one,
consistent with the results of mass assembly histories.
This indicates that the RDAGN hosts tend to become quenched $\sim$0.4-0.6\,Gyr earlier.

\subsection{Emission line excess in the nuclei of radio AGN hosts}\label{sec:spec}

Previous analyses have shown relatively marginal differences
in star formation properties between RDAGN hosts and their controls.
In this subsection,
we directly compare the observed spectra of RDAGN hosts and their controls,
focusing particularly on the emission lines.
Radio jets have long been known to ionize the ISM and create observable emission lines \citepads[e.g.,][]{2019ARA&A..57..467B}.
High-resolution jet simulations show that the kinetic energy in radio jets can be transferred to the surrounding gas by shocks in supersonic outflows \citepads{2017MNRAS.465.3291W}.

Most massive quiescent galaxies do not have strong emission line features.
To enhance the signal-to-noise ratio (S/N) of the emission lines, i.e., reduce the noise,
we leverage the abundant spectra from IFU observations and
stack (median) them in radial bins for both RDAGN hosts and their controls.
During the stacking process, all spectra are converted to the rest frame.
Spectra in each radial bin of a galaxy are first stacked by observed flux,
which ensures that each galaxy contributes equally to the population stack.
This approach highlights differences in emission and absorption lines between RDAGN hosts and controls.
In the differential spectra, stronger emission lines in RDAGN hosts appear as `emission',
while weaker lines appear as `absorption',
and vice versa for absorption lines.

\begin{figure}
\centering
\includegraphics[width=\hsize]{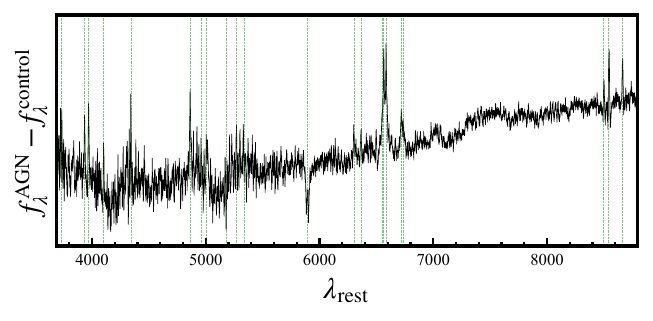}
\caption{
Illustration of the stacked differential spectrum between all RDAGN hosts and their controls from the nuclear regions ($\sim2.5 \arcsec$ diameter).
The spectrum is shown over the restframe wavelength range  3700$\AA$-8800$\AA$.
This wavelength range covers the most prominent emission and absorption lines from [\ion{O}{II}]$\lambda$3727 to \ion{Ca}{II}$\lambda$8662,
which are indicated by green dashed lines.
The `emission'- or `absorption'-like features indicate that those emission lines are stronger or weaker (respectively) in RDAGN hosts than in their control galaxy stacks.
The most prominent features of emission line excess are from 
[\ion{N}{II}]$\lambda$6550,6585 and H$\alpha$.
This kind of differential stacked spectra are created for different regions and different RDAGN populations, and analyzed in Section~\ref{sec:spec}.  
}
\label{fig:specall}%
\end{figure}

As an illustration, Fig.~\ref{fig:specall} shows the stacked differential spectra from the nuclear (within a $2.5 \arcsec$ diameter) regions of the radio AGN and their controls over the wavelength interval  3700$\AA$-8800$\AA$.
The strong optical emission and absorption lines are indicated in green lines.
In general, RDAGN hosts show redder spectra and several emission line excess or deficiency features in their nuclear regions.
The most prominent features are the [\ion{N}{II}]$\lambda$6550,6585 and H$\alpha$ emission line excess.
This excess is also confirmed by other less prominent lines, such as H$\beta$ and [\ion{S}{II}]$\lambda$6718,6733 doublets.
For absorption lines, RDAGN hosts have clearly weaker Ca lines
(CaH$\lambda$3934, CaK$\lambda$3968, and \ion{Ca}{II}$\lambda$8498,8542,8662),
appearing as emission-like features.
In contrast, the Mg $b$ and Na D indices show absorption-like features, indicating RDAGN hosts have lower Na and Mg abundance.
These absorption line features are not strong enough to accurately constrain the metallicity differences, so we only model the [\ion{N}{II}] and H$\alpha$ features in this work.

To investigate which population of  RDAGN are contributing to the stacked emission line excess and deficiency,
we divide the RDAGN sample into different subsamples by
jet luminosity, radio morphology, and stellar mass, respectively.
In each category, we create 16 stacked different spectra from these different subsamples and from four radial bins (0-0.5\,$R_{e}$, 0.5-1\,$R_{e}$, 1-1.5\,$R_{e}$,  $>1.5\,R_{e}$).
We then focus on the wavelength region between 6380$\AA$ and 6670$\AA$ to quantify the width and strength of the H$\alpha$+[\ion{N}{II}] features.
We subtract the continuum by fitting it polynomially
and then use the single component Gaussian model to fit the equivalent width and dispersion of these three lines.

In order to ensure that our spectral stacks of radio AGN are fully complete, we use the 90\% completeness cut in radio luminosity as a function of redshift to divide the radio AGN sample into four luminosity bins constrained by redshift. The median AGN luminosities for the bins are chosen to be $10^{22}$, $10^{23}$, $10^{24}$, and $10^{25}$ W Hz$^{-1}$. In total, 388 RDAGN are left to be used for the spectral stacking analysis.

\begin{figure}
\centering
\includegraphics[width=\hsize]{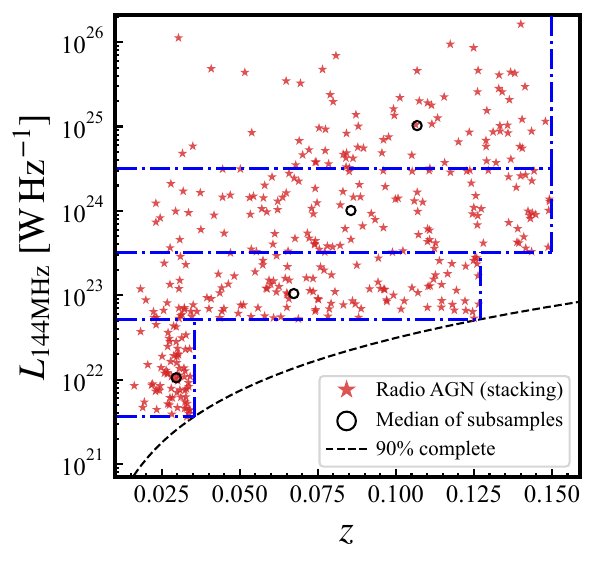}
  \caption{The radio luminosity completeness cut for the RDAGN sample used for spectral stacking. In different radio luminosity bins, the redshift limits of the sample are defined by the 90\% completeness cut. The median luminosities and redshifts of each subsample are plotted as black circles.}
     \label{fig:zcut}
\end{figure} 

\begin{figure*}
\centering
\includegraphics[width=\hsize]{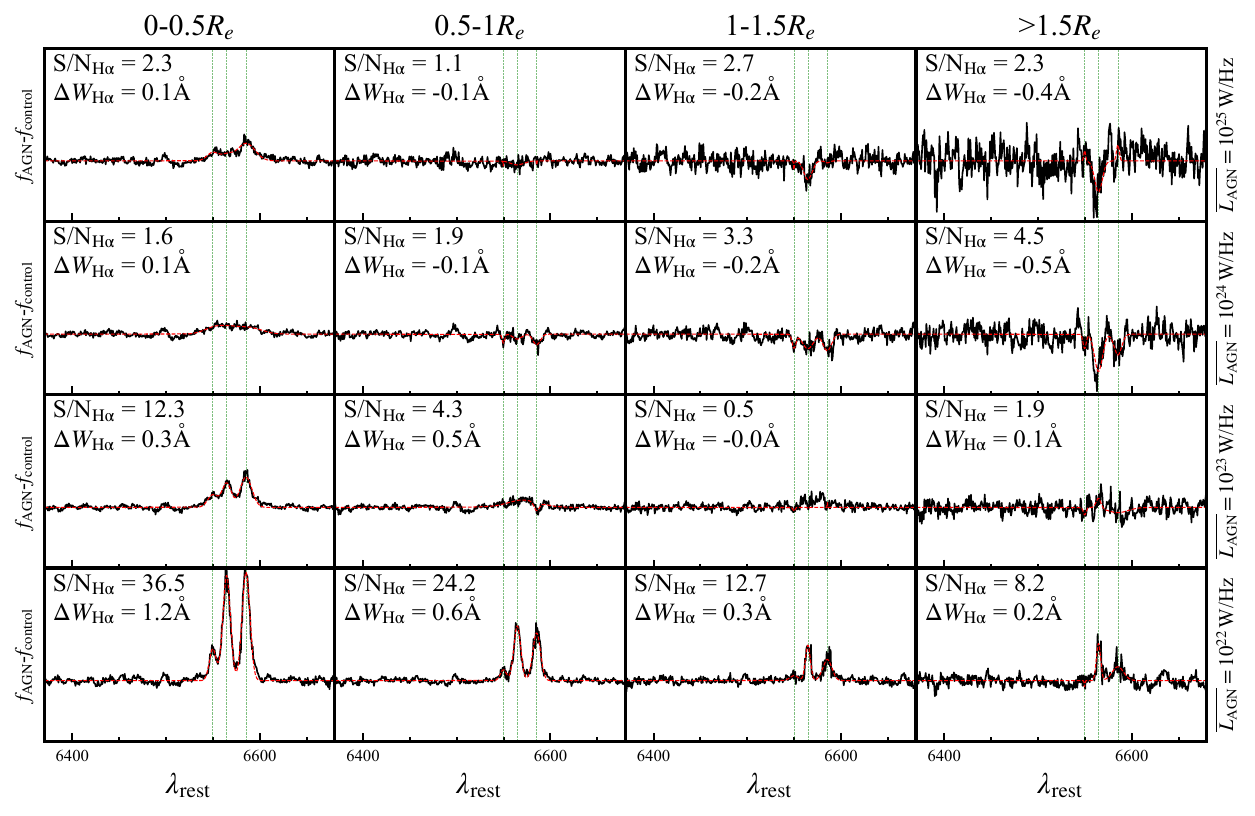}
\caption{
The stacked differential spectra between RDAGN hosts and their controls and the fitting results, are shown in the [\ion{N}{II}] doublets + H$\alpha$ (green dashed lines) wavelength range.
Stacking is done in subsamples with four $L_{\rm AGN}$ bins (powerful to weak from top to bottom) and in four radial bins (center to outskirts from left to right).
Red lines are the fitting results for the emission lines.
During the fitting, we assume that the three emission lines have individual Gaussian profiles.
The equivalent width of the H$\alpha$ and its signal-to-noise ratio derived from the fitting results are listed in each panel.
RDAGN hosts, especially the low luminosity ones, have emission line excess in their nuclear regions (the leftmost panels).
High $L_{\rm AGN}$ RDAGN hosts show weak emission line deficiency in their outskirts (right panels).
}
\label{fig:spec_diff}
\end{figure*}

Fig.~\ref{fig:spec_diff} illustrates this approach when we stack the spectra from RDAGN with different $L_{\rm AGN}$.
We set luminosity bins to make the median $L_{\rm AGN}$ values equal to $10^{22}, 10^{23}, 10^{24}$, and $\rm 10^{25}\ W\,Hz^{-1}$.
There are 70, 122, 121, and 75 RDAGN in these four luminosity bins, respectively.
Panels from top to bottom represent subsamples of four $L_{\rm AGN}$ bins, from powerful RDAGN to weak RDAGN.
Within each luminosity bin, the differential spectra are plotted for four radial bins
(0-0.5\,$R_{e}$, 0.5-1\,$R_{e}$, 1-1.5\,$R_{e}$,  $>1.5\,R_{e}$) from left to right.
We note that the outermost region, $r>1.5\,R_{e}$, is based on 
incomplete samples because of the field of view limit of MaNGA IFUs.
The equivalent width and the signal-to-noise ratio of H$\alpha$ are listed in each panel, where positive and negative values represent stronger or weaker H$\alpha$ in RDAGN hosts, respectively.
It can be seen that 
stacking can reveal equivalent width differences of less than 0.5$\AA$.

\begin{figure}
\centering
\includegraphics[width=\hsize]{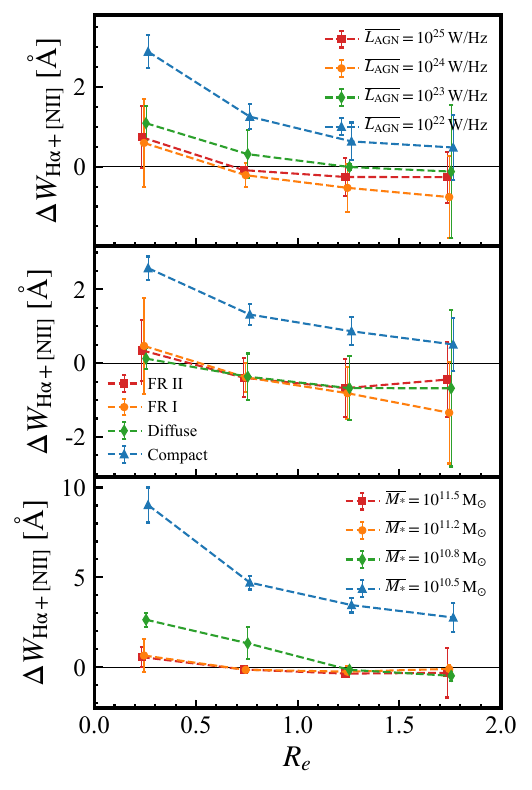}
  \caption{The equivalent width radial profile and the $3\sigma$ error-bars} of the $\rm H\alpha + [\ion{N}{II}]$ excess or deficiency ($\Delta W_{\rm H\alpha + [\ion{N}{II}]}$) in different RDAGN subsamples. Panels from top to bottom show the results when dividing RDAGN by jet luminosity, radio morphology, and stellar mass, respectively.
  Profiles of different subsamples are shown in different colors and markers. 
  Positive $\Delta W_{\rm H\alpha + [\ion{N}{II}]}$ represents that RDAGN have stronger emission lines and vice versa.
  Low luminosity ($\overline{L_{\rm AGN}} = \rm 10^{22}\ W\,Hz^{-1}$), compact ($<6\arcsec$), and lower mass RDAGN ($\overline{M_{*}} = \rm 10^{10.5}\, M_{\odot}$) contribute most to the nuclear excess.
  Weak emission line deficiency is found in the outskirts of large size RDAGN population.
     \label{fig:gradient}
\end{figure}

We use a similar approach for the other two categories,
radio morphology and stellar mass,
and summarize the equivalent width measurements in Fig~\ref{fig:gradient}.
The three panels from top to bottom show the results when dividing the RDAGN host sample
by $L_{\rm AGN}$, radio morphology, and stellar mass.
The calculation of equivalent width ($\Delta W_{\rm H\alpha + [\ion{N}{II}]}$) is based on the same method as in {\tt DAP} \citepads{2019AJ....158..160B}.
For radio morphology, we visually classify the RDAGN into FR\,I, FR\,II, diffuse, and compact (or unresolved) morphology based on the 6$\arcsec$ resolution LoTSS images.
The number of RDAGN in each morphology category is 81, 77, 38, and 192, respectively.
In the lowest panel, we divide the RDAGN hosts into four stellar mass bins
with median $M_*$ of $10^{10.5}, 10^{10.8}, 10^{11.2}$, and $\rm 10^{11.5} M_{\odot}$,
which include 47, 50, 145, and 146 galaxies, respectively.
Fig~\ref{fig:gradient} shows the radial profiles of the line excess.
Among different subsamples, the compact radio sources ($<6\arcsec$) show strong emission line excess, 
consistent with the scenario that the compact radio AGN are younger \citepads[e.g.][]{2024MNRAS.529.1472C} and therefore have a greater impact on gas ionization around the central SMBHs.
In addition, the lower mass RDAGN (median $M_{*}$: $\rm 10^{10.5}\, M_{\odot}$)  have the strongest H$\alpha$+[\ion{N}{II}] excess in their nuclear regions.
This indicates that the connection between radio AGN activity and optical emission is more common when the RDAGN hosts are less massive.
In the outskirts ($r > R_e$), we find weak emission line deficiency, mainly contributed by resolved RDAGN (FR\,I, FR\,II, and diffuse RDAGN).
These are likely to be older sources where the jet impacts the ISM at larger distances.

\begin{figure}
\centering
\includegraphics[width=\hsize]{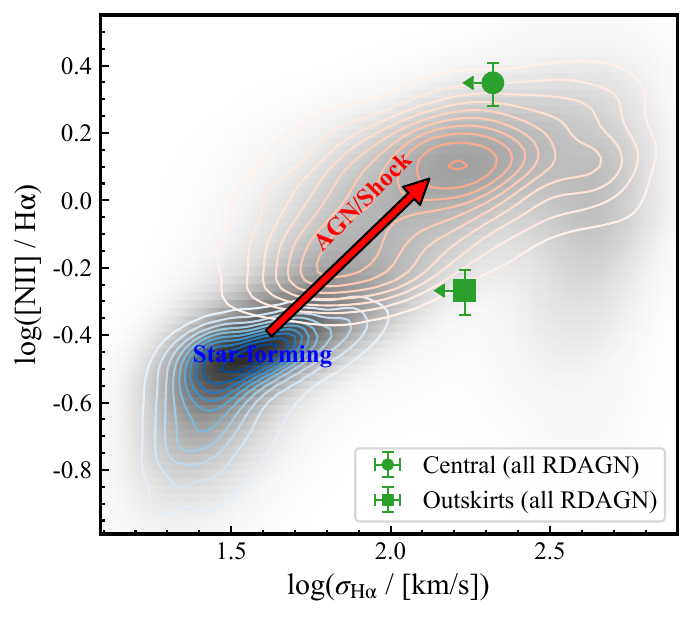}
  \caption{Ionization condition diagnostic of the emission line excess found in the nuclear regions of RDAGN hosts (green circle)
  as well as the deficiency found in the outskirts (green square).
  The green upper limits on $\sigma_{\rm H\alpha}$ are due to the unknown level of line broadening after stacking the spectra.
  The background gray-scale map is the distribution of the [\ion{N}{II}]/H$\alpha$ versus $\sigma_{\rm H\alpha}$ for the whole MaNGA sample.
  Star-forming-like emission and AGN or shock dominated emission regions as classified in the BPT diagram are plotted as blue and red contours, respectively.
  The red arrow indicates the direction in which the ionization becomes harder due to the AGN or shock contribution.
  In RDAGN hosts, the line ratio suggests that the nuclear emission line excess is created by AGN or shocks,
  while weaker star formation may be the reason for the emission line deficiency in the outskirts.
  }
     \label{fig:shock}
\end{figure} 

Excess emission lines in RDAGN hosts are only found in the innermost inner bins,
which indicates that the excess is likely related to the central AGN.
On the other hand, the deficiency is found in the outskirts at radii greater than a few kpcs, where the emission lines are not likely to be affected by the central AGN,
so this likely indicates a star formation deficiency in  the RDAGN hosts.
To examine this assumption, we examine the ionization condition of the ISM based on the velocity dispersion of H$\alpha$ ($\sigma_{\rm H\alpha}$) and the [\ion{N}{II}]/H$\alpha$ ratio.

In Fig~\ref{fig:shock}, we plot the distribution of the nuclear [\ion{N}{II}]/H$\alpha$ versus $\sigma_{\rm H\alpha}$ of all MaNGA galaxies using a logarithm scale (background gray contours).
Emission lines from \ion{H}{II} regions and AGN or shocks are marked by blue and red contours, which are used to classify galaxies according to location in the BPT diagram \citepads[e.g.,][]{2003MNRAS.346.1055K}.
The red arrow indicates the direction where AGN or shock ionization becomes more dominant.
We measure the the line ratio and $\sigma_{\rm H\alpha}$ of the emission excess for the whole RDAGN sample in the nuclear bin and the deficiency in the outskirts,
and plot them as a green circle and square respectively.
We note that the $\sigma_{\rm H\alpha}$ measurements are shown as upper limits, because the stacking procedure will bring an unknown level of line broadening.
The central line excess shows a high [\ion{N}{II}]/H$\alpha$ ratio and is located in the region predicated by AGN or shock models \citepads[e.g.,][]{2019MNRAS.487.4153D}. 
This supports the hypothesis  that the central RDAGN is heating and ionizing the nuclear gas.
However, the excess is weak ($\le 3\AA$) and not correlated with jet luminosity.
This suggests that the RDAGN have a weak and complex impact on the ISM properties in the central region of the galaxy.
It is worth noting that the resolution of the LoTSS images and MaNGA IFUs is at the kiloparsec level,
necessitating future high-resolution observations to confirm whether stronger effects are seen at smaller
distances from the SMBH.
In addition, high-resolution images are also key to distinguishing whether the emission is from the jet-related shock or from AGN photoionization.
Compared to the central region, the deficiency in the outskirts is located much closer to the star-forming region,
suggesting that the RDAGN hosts tend to be more quiescent in outer regions.

In summary, we find evidence of emission line excess in the nuclei of RDAGN host galaxies.
The excess is stronger in lower radio luminosity, more compact (smaller size of radio emission), and lower stellar mass RDAGN.
The line excess is likely linked to the interaction of the jet with the gas in the central regions of the host galaxies of  the RDAGN.
The line ratios suggest AGN photoionization or shock ionization,
and the excess is stronger towards the galaxy center.
We note that the level of emission line excess in our sample is not correlated with jet power in the range $10^{23}-10^{24}$ W Hz$^{-1}$. In the lowest luminosity bin, stronger excess is seen. 

We note that the sample is not large enough to split the sample into a bin of more than one quantity.
In order for the radio jet to  produce an emission line excess, the host galaxy must contain gas. 
Our radio-selected sample, however, includes many massive galaxies that lack gas to interact with the radio jets. 
This may explain the stronger line excess observed in lower mass RDAGN, 
as lower mass galaxies usually contain more gas \citepads[e.g.][]{2022ARA&A..60..319S}.
The excess observed in the lowest radio luminosity bin may also result from the fact that these are mainly found in low mass galaxies. 

\section{Discussion}\label{sec:discuss}
In this section, we discuss in more detail some of the main unsolved problems related to
the interpretation of the results in this paper and offer some ideas about how these may be
solved in the future.

\subsection{Selection effects}\label{sec:bias}
In studies focusing on the selection and analysis of RDAGN hosts,
addressing the potential biases introduced by the non-detections is crucial.
Simply defining the non-RDAGN sample based on the absence of radio detection or radio excess may lead to improper control samples.
For instance, a higher redshift radio non-detected galaxy might still have higher luminosity jets compared to a lower redshift weak RDAGN.
A SFG with an observed $L\rm _{144MHz}=10^{23}\,W\,Hz^{-1}$ without significant radio excess 
might conceal an $L\rm _{AGN}=10^{22}\,W\,Hz^{-1}$ AGN jet drowned in the emission from HII regions.
Both of these two galaxies are not appropriate to be selected as part of the  control sample of  $L\rm _{AGN}=10^{22}\,W\,Hz^{-1}$ RDAGN,
but they have been selected to be part of the control sample in most previous studies.
For the sensitivity of LoTSS, this effect is only negligible at $L\rm _{144MHz}>10^{24}\,W\,Hz^{-1}$, as shown in Fig.~\ref{fig:dsfms}.

In this work, we avoid this bias by calculating the $L\rm _{AGN}$ upper limits of all the non-detections and non-RDAGN.
The upper limits can help us to select control samples
that are certain to either contain no jets or, at worst, jets which are much fainter than those they are being controlled against.
Nonetheless, this approach may introduce bias against galaxies with higher SFRs.
For example, an $L\rm _{144MHz}=10^{23}\,W\,Hz^{-1}$ pure SFG without any AGN activity will not be selected as the control of an $L\rm _{AGN}=10^{23}\,W\,Hz^{-1}$ RDAGN with the same stellar mass.
Despite this bias against higher SFRs,
our control sample still exhibits SFRs about 0.2 dex higher than the RDAGN hosts, as also demonstrated in Section~\ref{sec:spec}.
In addition, the RDAGN hosts are dominated by massive galaxies, as shown in \ref{fig:sfms}, 
and the high SFR SFGs are rare at $M_*>10^{11}\rm \, M_{\odot}$.
These considerations ensure that our control sample selection is appropriately tailored
to mitigate selection biases and provide robust comparisons with RDAGN hosts.

\subsection{Disparity between the jet age and quenching time}

If the `radio-mode' AGN feedback can indeed suppress star formation, we would expect to observe a correlation between the time when the jet activity and the quenching occurred.
The timing of galaxies becoming quenched can be inferred from their mass assembly history.
On the other hand, estimating the age of radio jets is more complex and typically involves two popular methods: spectral age and dynamical age estimation.
The spectral age method relies on the shape of the synchrotron energy spectrum.
The synchrotron radiation in a fixed magnetic field will have energy losses that scale as $\partial E / \partial t \propto \nu^2$.
So if there is no source driving constant particle acceleration, 
the spectrum will become steeper because the flux density at high frequencies will decrease faster \citepads[e.g.,][]{2015MNRAS.454.3403H}.
The dynamical age is based on the size of a radio jet,
because the jet speed can be derived from the environment density \citepads[e.g.,][]{1991MNRAS.250..581F},
then the time scale can be calculated by the jet size and jet speed.
We note that these two ages are often observed to have large disparity \citepads[e.g.,][]{2013MNRAS.435.3353H}.

Most of the RDAGN in this study lack sufficient frequency coverage and spatial information
to accurately determine their spectral and dynamical jet ages.
However, even for extremely old jets with well-calibrated dynamical and spectral ages,
the jet ages are less than 1\,Gyr \citepads{2018MNRAS.474.3361T},
which is significantly shorter than the quiescent period observed in our RDAGN hosts.
As depicted in Fig.~\ref{fig:sfh}, RDAGN hosts are observed to have become quenched approximately 5 Gyr ago.
If both timescale measurements are accurate within factors of a few,
our results suggest that the onset of radio jet activity likely occurred after the host galaxy became quenched,
contradicting the scenario where AGN appear and then suppress star formation.

This then implies that the observed RDAGN activity is not responsible for the host quenching.
Several recent studies have argued that different AGN populations may represent different evolution phases of AGN,
and the radio AGN represent the very late stages of a cycle of AGN activity
\citepads[e.g.,][]{2019MNRAS.488.3109K,2024A&A...691A.124A}.
If so, this would explain why the radio jet activities may occur later than the quenching,
since the star formation suppression happened in AGN's early stages when the radiation feedback is strong.
Even with the short timescale of radio jet activity, we still find high RDAGN incidence in the most massive galaxies ($>$50\%, see Fig.~\ref{fig:sfms}),
which indicates that radio AGN are common in massive galaxies with a high duty cycle, and the radio mode might be an important ingredient in AGN activities.
However, it is still difficult to constrain the duty cycle and evolution timescale of AGN according to the observations.

\subsection{The triggering of radio AGN}
\begin{figure}
\centering
\includegraphics[width=\hsize]{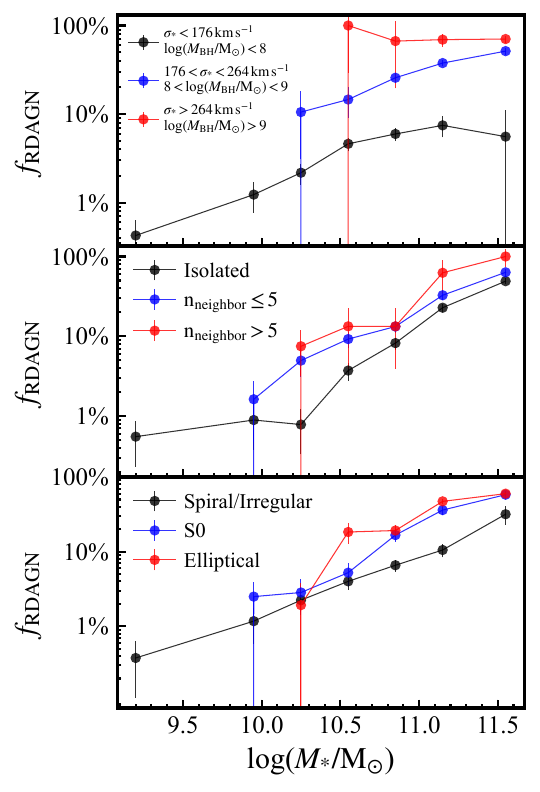}
  \caption{Incidence of RDAGN as a function of stellar mass, shown for different galaxy populations. From top to bottom, the galaxies are divided by stellar velocity dispersion (an indicator of the black hole mass), the number of neighbors within 200\,kpc (an indicator of the cosmic environment), and Hubble type (galaxy morphology), respectively.
  RDAGN tend to be located in host galaxies with larger stellar velocity dispersions, denser environments, and early-type morphologies.
  We find that most galaxies with the largest stellar dispersion ($\sigma_{*}>264\rm \, km\,s^{-1}$) host RDAGN, 
  which suggests that the most massive SMBHs are always accreting in the radio mode.
  }
     \label{fig:trigger}
\end{figure} 

In Figure 5, we showed that the fraction of galaxies that host radio-loud AGN is a very
strong function of the stellar mass of the galaxy. 
In this section, we attempt to pinpoint the underlying cause for why radio AGN activity is more common in massive galaxies by including
additional parameters in the analysis, namely
SMBH mass, the cosmic environment, and the galaxy morphology.
The SMBH mass is estimated by the $M_{\rm BH}-\sigma_{*}$ relation used in \citetads{2013ApJ...764..184M}.
The $\sigma_{*}$ is fitted from the MaNGA spectra using {\tt DAP} \citepads{2019AJ....158..231W}.
We adopt the numbers of neighbors within the projected 200 kpc radius and line-of-sight $\Delta v\rm <500\,km\,s^{-1}$ as an approximation for the environmental impact,
which are taken from the {\tt MaNGA-GEMA} catalog\footnote{\url{https://data.sdss.org/datamodel/files/MANGA_GEMA/GEMA_VER}} (Argudo-Fernádez et al. in prep).
We use the Hubble type visually classified by \citetads{2022MNRAS.512.2222V} and divided galaxies into elliptical, S0, and late-type (spiral and irregular galaxies).

We calculate the incidence of RDAGN as a function of stellar mass,
and plot the results for subsamples divided by the 3 additional parameters in Fig.\ref{fig:trigger}.
In the upper panel, the RDAGN incidence is clearly higher in high $\sigma_{*}$ systems
at fixed stellar mass. The middle and lower panels indicate that
at fixed stellar mass, RDAGN prefer denser cosmic environment and early-type galaxies,
which is consistent with  previous findings \citepads[e.g.,][]{2005MNRAS.362...25B}.  
The trends as a function of $\sigma_{*}$ at fixed stellar mass are by far the strongest ones,
suggesting that black hole mass is an important parameter, after stellar mass, that controls
the fraction of galaxies with radio mode AGN activity.
We note that at a similar Eddington ratio, the more massive black holes will have higher luminosity and thus are more likely to be detected.
\citetads{2019A&A...622A..17S} used a larger single-fiber observed sample and showed that the RDAGN incidence dependence on $M_{\rm BH}$ is marginal.
This inconsistency is partly due to that they made a radio luminosity cut on the radio AGN sample.
After applying a similar cut ($L\rm_{150MHz}>10^{22.5}\ W\,Hz^{-1}$), our fraction values generally agree with the results in \citetads{2019A&A...622A..17S} ($f_{\rm RDAGN}\sim5\% - 10\%$ at $M_{*}\sim 10^{11} \rm M_{\odot}$).
However, the dependence of the RDAGN incidence on $M_{\rm BH}$ still exists.
Possible reasons could be our much smaller sample size, and/or the different redshift range ($z<0.15$ versus $z<0.3$).

\subsection{Implications for the AGN feedback scenario}

Combining all the observation results found in this paper,
we discuss some possible new implications for the AGN feedback scenario.
We note that this discussion focuses on the moderate and faint radio AGN, 
where a direct connection between the radio AGN and the emission-line properties of the ISM is clearly seen.
The more luminous radio jets may dump most of their energy at large distances from the galaxy, and the link to quenching effects may not be as apparent.

Because radio AGN are located in denser environments, at fixed mass, they
are likely to have assembled  earlier and faster.
The observed higher stellar dispersion suggests that their central SMBHs are more massive, 
indicating possible stronger AGN activity in the past.
This past AGN activity 
may have  played an important role in quenching the host galaxies through gas heating, fast consumption
of the gas, or gas outflows \citepads[e.g.,][]{2024ApJ...961L..39S}.
The early-type morphology and low angular momentum \citepads{2023A&A...673A..12Z} indicates past galaxy-galaxy merger events, which may trigger
the past AGN activity and also the fast mass growth, as well as the morphological transition
from disk to spheroid \citepads[e.g.][]{2004AJ....128.2098R}.
This is also consistent with some simulation results which suggest that episodic AGN events, rather than one time AGN event, contribute to the final quenching of the host galaxies \citepads[e.g.][]{2022MNRAS.512.1052P,2024Galax..12...17H}.

At a later stage, the galaxy becomes quiescent due to the lack of cold gas fueling. 
This also prevents the central SMBH from accreting efficiently and radiatively, 
and the current most massive SMBHs tend to appear as radio-mode AGN.
In our sample,
the RDAGN hosts are dominated by quiescent galaxies and show weaker star formation activities in the outskirts,
which suggests that the radio AGN may help the host galaxies to maintain quiescence.
The AGN- or shock like emission excess found in the nuclear regions of RDAGN hosts indicates that radio AGN ionize and heat the surrounding gas. 
One remaining mystery is exactly how the
weak and compact jets associated with the central excess phenomenon can maintain the quiescence of the host galaxies globally.

Finally, although a large-scale connection between radio AGN and the heating of surrounding gas has been observed in many central galaxies of massive clusters \citepads[e.g.,][]{2012ARA&A..50..455F},
we still need more observational evidence of the jet-gas connection in faint radio AGN on galaxy scales to obtain a full view of such physical processes in different galaxy environments and populations.

\subsection{Future directions}

This work is based on the radio images and optical IFUs with spatial resolutions on the order of arcseconds.
In the future, new telescopes and the update of current facilities 
can carry out large radio sky surveys with sub-arcsec resolution, e.g., The International LOFAR Two-metre Sky Survey\footnote{\url{https://lofar-surveys.org/ilotss.html}}, 
the Next Generation Very Large Array \citepads[ngVLA,][]{2019clrp.2020...32D}, 
the Square Kilometre Array\footnote{\url{https://www.skao.int/}} (SKA), etc..
At low redshifts, they can resolve the jet features on physical scales of a few hundred of parsecs.
As discussed in Section~\ref{sec:global}, the higher resolution is also essential to decompose the faint radio population in AGN and star-forming galaxies. 
This can help
us to understand the radio AGN luminosity function in the faint end.
The wide frequency coverage of different instruments can also help us build the full synchrotron SED of radio AGN,
which is essential for constraining jet ages.

Meanwhile, there are more and more deep and high-resolution IFU samples.
Future high-resolution IFUs can provide emission line maps around the AGN at parsec-scale resolution.
Combined with high-resolution radio images, the morphology of the radio continuum and emission lines may reveal the detailed process of how the jets from AGN interact with the ISM.
In addition to advancements in radio and optical observations, future large-scale surveys at other wavelengths, 
particularly in the X-ray and infrared, are expected to match the resolution and
sensitivity achieved at optical and radio bands.
This multi-wavelength approach will provide comprehensive information about the different component structures of AGN,
allowing for a more complete understanding of the structure and the evolution of AGN.

\section{Summary}\label{sec:conclusion}

We build the largest optical IFU - radio continuum sample to date 
consisting of 5548 galaxies by matching the MaNGA survey with LoTSS-DR2 observations.
We revisit the tight linear SFR - radio luminosity relation using IFU data and find an intrinsic error of $\sim0.23$\,dex.
Using this relation, we identify 616 radio AGN with excessive radio emission, and calculate their jet luminosities.
We analyze their global star formation rates and their correlation with jet luminosity.
We then derive the radially resolved  mass assembly history by fitting the IFU spectra in radial bins
and comparing the difference between radio AGN hosts and their control samples
We also stack the spectra in different radial bins to 
investigate the impact of radio AGN on the emission lines of their host galaxies.
Our results can be summarized as follows:

\begin{enumerate}
    \item The linear relation in the SFR and $L_{\rm 144MHz}$ found by LoTSS-MaNGA sample is tight and can be described by Equation~\ref{eq:l144-sfr}:
    $
   {\rm log}(\frac{L_{\rm 144MHz}}{\rm W\,Hz^{-1}})=1.16\times {\rm log}(\frac{\rm SFR}{\rm M_{\odot}\,yr^{-1}}) + 21.99 
    $
    , which is in very good agreement with previous results based on non-IFU samples.
    \item At least 41\% $M_* \rm > 10^{11}\,M_{\odot}$ galaxies are found to host a radio AGN.
    Radio AGN host galaxies are dominated by massive quiescent galaxies, i.e., 92\% lie below the $-3\sigma$ of star formation main sequence.
    The quenching level, $\Delta$SFR, is not related to the jet luminosity (Pearson correlation coefficient $\rho \sim -0.1$).
    \item The mass assembly history of radio AGN hosts follows an `inside-out' scenario, similar to normal massive galaxies.
    Compared to a mass-matched control sample, radio AGN hosts grow faster at all radii, especially in their central regions.
    They grew to 90 percent of current mass at lookback times of  4-6\,Gyr, and 0.4-0.6\,Gyr earlier than their controls.
    The mass assembly history reconstructed from the optical spectra is consistent with that in simulations incorporating the radio-mode AGN feedback model.
    \item Stacked spectra comparisons reveal emission line excess exclusively in the central region of radio AGN hosts, especially in low mass and radio-compact RDAGN.
    The high ratio of the excessive [\ion{N}{II}] and H$\alpha$ supports the scenario that the emission line excess is linked to AGN or shock ionization, but it remains to be understood why the excess is not dependent upon jet luminosity.
\end{enumerate}

Considering that the timescale of jet activities is usually less than 1\,Gyr,
the quenching of their host galaxies that happened several Gyrs ago was likely not due to the current RDAGN.
However, episodic cycles of radio AGN may play an important role in maintaining the global quiescence of the host galaxies and preventing the most massive galaxies from further growth. This conjecture is supported by the 
high fraction of quiescent galaxies in RDAGN hosts,
as well as the high incidence of RDAGN in massive galaxies.
The lack of a clear correlation between jet luminosity and feedback features suggests that
energy transfer from radio jets to the ISM is an inefficient and complex process.
Exactly how radio jets affect the star formation and interact with the gas in galaxies is yet to be fully understood.
Future studies with larger samples and higher spatial resolution observations 
at different wavelengths are expected to shed more light on the
detailed physical processes underlying `radio-mode' AGN feedback.

\begin{acknowledgements}

    PNB is grateful for support from the UK STFC via grants ST/V000594/1 and ST/Y000951/1.
    SS is grateful for support from the UK Science and Technology Facilities Council (STFC) via grant ST/X508408/1.
    
    Funding for the Sloan Digital Sky 
    Survey IV has been provided by the 
    Alfred P. Sloan Foundation, the U.S. 
    Department of Energy Office of 
    Science, and the Participating 
    Institutions. 
    
    SDSS-IV acknowledges support and 
    resources from the Center for High 
    Performance Computing  at the 
    University of Utah. The SDSS 
    website is www.sdss4.org.
    
    SDSS-IV is managed by the 
    Astrophysical Research Consortium 
    for the Participating Institutions 
    of the SDSS Collaboration including 
    the Brazilian Participation Group, 
    the Carnegie Institution for Science, 
    Carnegie Mellon University, Center for 
    Astrophysics | Harvard \& 
    Smithsonian, the Chilean Participation 
    Group, the French Participation Group, 
    Instituto de Astrof\'isica de 
    Canarias, The Johns Hopkins 
    University, Kavli Institute for the 
    Physics and Mathematics of the 
    Universe (IPMU) / University of 
    Tokyo, the Korean Participation Group, 
    Lawrence Berkeley National Laboratory, 
    Leibniz Institut f\"ur Astrophysik 
    Potsdam (AIP),  Max-Planck-Institut 
    f\"ur Astronomie (MPIA Heidelberg), 
    Max-Planck-Institut f\"ur 
    Astrophysik (MPA Garching), 
    Max-Planck-Institut f\"ur 
    Extraterrestrische Physik (MPE), 
    National Astronomical Observatories of 
    China, New Mexico State University, 
    New York University, University of 
    Notre Dame, Observat\'ario 
    Nacional / MCTI, The Ohio State 
    University, Pennsylvania State 
    University, Shanghai 
    Astronomical Observatory, United 
    Kingdom Participation Group, 
    Universidad Nacional Aut\'onoma 
    de M\'exico, University of Arizona, 
    University of Colorado Boulder, 
    University of Oxford, University of 
    Portsmouth, University of Utah, 
    University of Virginia, University 
    of Washington, University of 
    Wisconsin, Vanderbilt University, 
    and Yale University.
    
    LOFAR \citepads{2013A&A...556A...2V}
    is the Low Frequency Array designed and constructed by 
    ASTRON. It has observing, data processing, and data storage facilities in several countries, 
    which are owned by various parties (each with their own funding sources), and that are 
    collectively operated by the ILT foundation under a joint scientific policy. The ILT resources 
    have benefited from the following recent major funding sources: CNRS-INSU, Observatoire de 
    Paris and Université d'Orléans, France; BMBF, MIWF-NRW, MPG, Germany; Science 
    Foundation Ireland (SFI), Department of Business, Enterprise and Innovation (DBEI), Ireland; 
    NWO, The Netherlands; The Science and Technology Facilities Council, UK; Ministry of 
    Science and Higher Education, Poland; The Istituto Nazionale di Astrofisica (INAF), Italy. 
    This research made use of the Dutch national e-infrastructure with support of the SURF 
    Cooperative (e-infra 180169) and the LOFAR e-infra group. The Jülich LOFAR Long Term 
    Archive and the German LOFAR network are both coordinated and operated by the Jülich 
    Supercomputing Centre (JSC), and computing resources on the supercomputer JUWELS at JSC 
    were provided by the Gauss Centre for Supercomputing e.V. (grant CHTB00) through the John 
    von Neumann Institute for Computing (NIC). 
    This research made use of the University of Hertfordshire high-performance computing facility
    and the LOFAR-UK computing facility located at the University of Hertfordshire and supported 
    by STFC [ST/P000096/1], and of the Italian LOFAR IT computing infrastructure supported and 
    operated by INAF, and by the Physics Department of Turin university (under an agreement with 
    Consorzio Interuniversitario per la Fisica Spaziale) at the C3S Supercomputing Centre, Italy.
      
    Part of this work was supported by the German
    \emph{Deut\-sche For\-schungs\-ge\-mein\-schaft, DFG\/} project
    number Ts\~17/2--1.
\end{acknowledgements}

%
\bibliographystyle{aa} 
\bibliography{article} 

\begin{thebibliography}{115}
\expandafter\ifx\csname natexlab\endcsname\relax\def\natexlab#1{#1}\fi

\bibitem[{{Abdurro'uf} {et~al.}(2022){Abdurro'uf}, {Accetta}, {Aerts}, {Silva
  Aguirre}, {Ahumada}, {Ajgaonkar}, {Filiz Ak}, {Alam}, {Allende Prieto},
  {Almeida}, {Anders}, {Anderson}, {Andrews}, {Anguiano}, {Aquino-Ort{\'\i}z},
  {Arag{\'o}n-Salamanca}, {Argudo-Fern{\'a}ndez}, {Ata}, {Aubert},
  {Avila-Reese}, {Badenes}, {Barb{\'a}}, {Barger}, {Barrera-Ballesteros},
  {Beaton}, {Beers}, {Belfiore}, {Bender}, {Bernardi}, {Bershady}, {Beutler},
  {Bidin}, {Bird}, {Bizyaev}, {Blanc}, {Blanton}, {Boardman}, {Bolton},
  {Boquien}, {Borissova}, {Bovy}, {Brandt}, {Brown}, {Brownstein}, {Brusa},
  {Buchner}, {Bundy}, {Burchett}, {Bureau}, {Burgasser}, {Cabang}, {Campbell},
  {Cappellari}, {Carlberg}, {Wanderley}, {Carrera}, {Cash}, {Chen}, {Chen},
  {Cherinka}, {Chiappini}, {Choi}, {Chojnowski}, {Chung}, {Clerc}, {Cohen},
  {Comerford}, {Comparat}, {da Costa}, {Covey}, {Crane}, {Cruz-Gonzalez},
  {Culhane}, {Cunha}, {Dai}, {Damke}, {Darling}, {Davidson}, {Davies},
  {Dawson}, {De Lee}, {Diamond-Stanic}, {Cano-D{\'\i}az}, {S{\'a}nchez},
  {Donor}, {Duckworth}, {Dwelly}, {Eisenstein}, {Elsworth}, {Emsellem},
  {Eracleous}, {Escoffier}, {Fan}, {Farr}, {Feng}, {Fern{\'a}ndez-Trincado},
  {Feuillet}, {Filipp}, {Fillingham}, {Frinchaboy}, {Fromenteau}, {Galbany},
  {Garc{\'\i}a}, {Garc{\'\i}a-Hern{\'a}ndez}, {Ge}, {Geisler}, {Gelfand},
  {G{\'e}ron}, {Gibson}, {Goddy}, {Godoy-Rivera}, {Grabowski}, {Green},
  {Greener}, {Grier}, {Griffith}, {Guo}, {Guy}, {Hadjara}, {Harding},
  {Hasselquist}, {Hayes}, {Hearty}, {Hern{\'a}ndez}, {Hill}, {Hogg},
  {Holtzman}, {Horta}, {Hsieh}, {Hsu}, {Hsu}, {Huber}, {Huertas-Company},
  {Hutchinson}, {Hwang}, {Ibarra-Medel}, {Chitham}, {Ilha}, {Imig}, {Jaekle},
  {Jayasinghe}, {Ji}, {Johnson}, {Jones}, {J{\"o}nsson}, {Katkov}, {Khalatyan},
  {Kinemuchi}, {Kisku}, {Knapen}, {Kneib}, {Kollmeier}, {Kong}, {Kounkel},
  {Kreckel}, {Krishnarao}, {Lacerna}, {Lane}, {Langgin}, {Lavender}, {Law},
  {Lazarz}, {Leung}, {Leung}, {Lewis}, {Li}, {Li}, {Lian}, {Liang}, {Lin},
  {Lin}, {Lin}, {Lintott}, {Long}, {Longa-Pe{\~n}a}, {L{\'o}pez-Cob{\'a}},
  {Lu}, {Lundgren}, {Luo}, {Mackereth}, {de la Macorra}, {Mahadevan},
  {Majewski}, {Manchado}, {Mandeville}, {Maraston}, {Margalef-Bentabol},
  {Masseron}, {Masters}, {Mathur}, {McDermid}, {Mckay}, {Merloni},
  {Merrifield}, {Meszaros}, {Miglio}, {Di Mille}, {Minniti}, {Minsley},
  {Monachesi}, {Moon}, {Mosser}, {Mulchaey}, {Muna}, {Mu{\~n}oz}, {Myers},
  {Myers}, {Nadathur}, {Nair}, {Nandra}, {Neumann}, {Newman}, {Nidever},
  {Nikakhtar}, {Nitschelm}, {O'Connell}, {Garma-Oehmichen}, {Luan Souza de
  Oliveira}, {Olney}, {Oravetz}, {Ortigoza-Urdaneta}, {Osorio}, {Otter},
  {Pace}, {Padilla}, {Pan}, {Pan}, {Parikh}, {Parker}, {Peirani}, {Pe{\~n}a
  Ram{\'\i}rez}, {Penny}, {Percival}, {Perez-Fournon}, {Pinsonneault},
  {Poidevin}, {Poovelil}, {Price-Whelan}, {B{\'a}rbara de Andrade Queiroz},
  {Raddick}, {Ray}, {Rembold}, {Riddle}, {Riffel}, {Riffel}, {Rix}, {Robin},
  {Rodr{\'\i}guez-Puebla}, {Roman-Lopes}, {Rom{\'a}n-Z{\'u}{\~n}iga}, {Rose},
  {Ross}, {Rossi}, {Rubin}, {Salvato}, {S{\'a}nchez}, {S{\'a}nchez-Gallego},
  {Sanderson}, {Santana Rojas}, {Sarceno}, {Sarmiento}, {Sayres}, {Sazonova},
  {Schaefer}, {Schiavon}, {Schlegel}, {Schneider}, {Schultheis}, {Schwope},
  {Serenelli}, {Serna}, {Shao}, {Shapiro}, {Sharma}, {Shen}, {Shetrone}, {Shu},
  {Simon}, {Skrutskie}, {Smethurst}, {Smith}, {Sobeck}, {Spoo}, {Sprague},
  {Stark}, {Stassun}, {Steinmetz}, {Stello}, {Stone-Martinez},
  {Storchi-Bergmann}, {Stringfellow}, {Stutz}, {Su}, {Taghizadeh-Popp},
  {Talbot}, {Tayar}, {Telles}, {Teske}, {Thakar}, {Theissen}, {Tkachenko},
  {Thomas}, {Tojeiro}, {Hernandez Toledo}, {Troup}, {Trump}, {Trussler},
  {Turner}, {Tuttle}, {Unda-Sanzana}, {V{\'a}zquez-Mata}, {Valentini},
  {Valenzuela}, {Vargas-Gonz{\'a}lez}, {Vargas-Maga{\~n}a}, {Alfaro},
  {Villanova}, {Vincenzo}, {Wake}, {Warfield}, {Washington}, {Weaver},
  {Weijmans}, {Weinberg}, {Weiss}, {Westfall}, {Wild}, {Wilde}, {Wilson},
  {Wilson}, {Wilson}, {Wolf}, {Wood-Vasey}, {Yan}, {Zamora}, {Zasowski},
  {Zhang}, {Zhao}, {Zheng}, {Zheng}, \& {Zhu}}]{2022ApJS..259...35A}
{Abdurro'uf}, {Accetta}, K., {Aerts}, C., {et~al.} 2022, \apjs, 259, 35

\bibitem[{{Alb{\'a}n} {et~al.}(2024){Alb{\'a}n}, {Wylezalek}, {Comerford},
  {Greene}, \& {Riffel}}]{2024A&A...691A.124A}
{Alb{\'a}n}, M., {Wylezalek}, D., {Comerford}, J.~M., {Greene}, J.~E., \&
  {Riffel}, R.~A. 2024, \aap, 691, A124

\bibitem[{{Baldwin}(1997)}]{1997ASPC..113...80B}
{Baldwin}, J.~A. 1997, in Astronomical Society of the Pacific Conference
  Series, Vol. 113, IAU Colloq. 159: Emission Lines in Active Galaxies: New
  Methods and Techniques, ed. B.~M. {Peterson}, F.-Z. {Cheng}, \& A.~S.
  {Wilson}, 80

\bibitem[{{Baldwin} {et~al.}(1981){Baldwin}, {Phillips}, \&
  {Terlevich}}]{1981PASP...93....5B}
{Baldwin}, J.~A., {Phillips}, M.~M., \& {Terlevich}, R. 1981, \pasp, 93, 5

\bibitem[{{Barvainis}(1987)}]{1987ApJ...320..537B}
{Barvainis}, R. 1987, \apj, 320, 537

\bibitem[{{Becker} {et~al.}(1995){Becker}, {White}, \&
  {Helfand}}]{1995ApJ...450..559B}
{Becker}, R.~H., {White}, R.~L., \& {Helfand}, D.~J. 1995, \apj, 450, 559

\bibitem[{{Begelman} {et~al.}(1984){Begelman}, {Blandford}, \&
  {Rees}}]{1984RvMP...56..255B}
{Begelman}, M.~C., {Blandford}, R.~D., \& {Rees}, M.~J. 1984, Reviews of Modern
  Physics, 56, 255

\bibitem[{{Belfiore} {et~al.}(2019){Belfiore}, {Westfall}, {Schaefer},
  {Cappellari}, {Ji}, {Bershady}, {Tremonti}, {Law}, {Yan}, {Bundy}, {Shetty},
  {Drory}, {Thomas}, {Emsellem}, \& {S{\'a}nchez}}]{2019AJ....158..160B}
{Belfiore}, F., {Westfall}, K.~B., {Schaefer}, A., {et~al.} 2019, \aj, 158, 160

\bibitem[{{Best} \& {Heckman}(2012)}]{2012MNRAS.421.1569B}
{Best}, P.~N. \& {Heckman}, T.~M. 2012, \mnras, 421, 1569

\bibitem[{{Best} {et~al.}(2005{\natexlab{a}}){Best}, {Kauffmann}, {Heckman},
  {Brinchmann}, {Charlot}, {Ivezi{\'c}}, \& {White}}]{2005MNRAS.362...25B}
{Best}, P.~N., {Kauffmann}, G., {Heckman}, T.~M., {et~al.} 2005{\natexlab{a}},
  \mnras, 362, 25

\bibitem[{{Best} {et~al.}(2005{\natexlab{b}}){Best}, {Kauffmann}, {Heckman}, \&
  {Ivezi{\'c}}}]{2005MNRAS.362....9B}
{Best}, P.~N., {Kauffmann}, G., {Heckman}, T.~M., \& {Ivezi{\'c}}, {\v{Z}}.
  2005{\natexlab{b}}, \mnras, 362, 9

\bibitem[{{Best} {et~al.}(2023){Best}, {Kondapally}, {Williams}, {Cochrane},
  {Duncan}, {Hale}, {Haskell}, {Ma{\l}ek}, {McCheyne}, {Smith}, {Wang},
  {Botteon}, {Bonato}, {Bondi}, {Calistro Rivera}, {Gao}, {G{\"u}rkan},
  {Hardcastle}, {Jarvis}, {Mingo}, {Miraghaei}, {Morabito}, {Nisbet},
  {Prandoni}, {R{\"o}ttgering}, {Sabater}, {Shimwell}, {Tasse}, \& {van
  Weeren}}]{2023MNRAS.523.1729B}
{Best}, P.~N., {Kondapally}, R., {Williams}, W.~L., {et~al.} 2023, \mnras, 523,
  1729

\bibitem[{{Biermann}(1976)}]{1976A&A....53..295B}
{Biermann}, P. 1976, \aap, 53, 295

\bibitem[{{Blandford} {et~al.}(2019){Blandford}, {Meier}, \&
  {Readhead}}]{2019ARA&A..57..467B}
{Blandford}, R., {Meier}, D., \& {Readhead}, A. 2019, \araa, 57, 467

\bibitem[{{Blanton} {et~al.}(2011){Blanton}, {Kazin}, {Muna}, {Weaver}, \&
  {Price-Whelan}}]{2011AJ....142...31B}
{Blanton}, M.~R., {Kazin}, E., {Muna}, D., {Weaver}, B.~A., \& {Price-Whelan},
  A. 2011, \aj, 142, 31

\bibitem[{{Blanton} \& {Roweis}(2007)}]{2007AJ....133..734B}
{Blanton}, M.~R. \& {Roweis}, S. 2007, \aj, 133, 734

\bibitem[{{Brown} {et~al.}(2017){Brown}, {Moustakas}, {Kennicutt}, {Bonne},
  {Intema}, {de Gasperin}, {Boquien}, {Jarrett}, {Cluver}, {Smith}, {da Cunha},
  {Imanishi}, {Armus}, {Brandl}, \& {Peek}}]{2017ApJ...847..136B}
{Brown}, M. J.~I., {Moustakas}, J., {Kennicutt}, R.~C., {et~al.} 2017, \apj,
  847, 136

\bibitem[{{Bundy} {et~al.}(2015){Bundy}, {Bershady}, {Law}, {Yan}, {Drory},
  {MacDonald}, {Wake}, {Cherinka}, {S{\'a}nchez-Gallego}, {Weijmans}, {Thomas},
  {Tremonti}, {Masters}, {Coccato}, {Diamond-Stanic}, {Arag{\'o}n-Salamanca},
  {Avila-Reese}, {Badenes}, {Falc{\'o}n-Barroso}, {Belfiore}, {Bizyaev},
  {Blanc}, {Bland-Hawthorn}, {Blanton}, {Brownstein}, {Byler}, {Cappellari},
  {Conroy}, {Dutton}, {Emsellem}, {Etherington}, {Frinchaboy}, {Fu}, {Gunn},
  {Harding}, {Johnston}, {Kauffmann}, {Kinemuchi}, {Klaene}, {Knapen},
  {Leauthaud}, {Li}, {Lin}, {Maiolino}, {Malanushenko}, {Malanushenko}, {Mao},
  {Maraston}, {McDermid}, {Merrifield}, {Nichol}, {Oravetz}, {Pan}, {Parejko},
  {Sanchez}, {Schlegel}, {Simmons}, {Steele}, {Steinmetz}, {Thanjavur},
  {Thompson}, {Tinker}, {van den Bosch}, {Westfall}, {Wilkinson}, {Wright},
  {Xiao}, \& {Zhang}}]{2015ApJ...798....7B}
{Bundy}, K., {Bershady}, M.~A., {Law}, D.~R., {et~al.} 2015, \apj, 798, 7

\bibitem[{{Calzetti} {et~al.}(2000){Calzetti}, {Armus}, {Bohlin}, {Kinney},
  {Koornneef}, \& {Storchi-Bergmann}}]{2000ApJ...533..682C}
{Calzetti}, D., {Armus}, L., {Bohlin}, R.~C., {et~al.} 2000, \apj, 533, 682

\bibitem[{{Chabrier}(2003)}]{2003PASP..115..763C}
{Chabrier}, G. 2003, \pasp, 115, 763

\bibitem[{{Chi} \& {Wolfendale}(1990)}]{1990MNRAS.245..101C}
{Chi}, X. \& {Wolfendale}, A.~W. 1990, \mnras, 245, 101

\bibitem[{{Chilufya} {et~al.}(2024){Chilufya}, {Hardcastle}, {Pierce},
  {Croston}, {Mingo}, {Zheng}, {Baldi}, \&
  {R{\"o}ttgering}}]{2024MNRAS.529.1472C}
{Chilufya}, J., {Hardcastle}, M.~J., {Pierce}, J.~C.~S., {et~al.} 2024, \mnras,
  529, 1472

\bibitem[{{Cid Fernandes} {et~al.}(2011){Cid Fernandes}, {Stasi{\'n}ska},
  {Mateus}, \& {Vale Asari}}]{2011MNRAS.413.1687C}
{Cid Fernandes}, R., {Stasi{\'n}ska}, G., {Mateus}, A., \& {Vale Asari}, N.
  2011, \mnras, 413, 1687

\bibitem[{{Cochrane} {et~al.}(2023){Cochrane}, {Kondapally}, {Best}, {Sabater},
  {Duncan}, {Smith}, {Hardcastle}, {R{\"o}ttgering}, {Prandoni}, {Haskell},
  {G{\"u}rkan}, \& {Miley}}]{2023MNRAS.523.6082C}
{Cochrane}, R.~K., {Kondapally}, R., {Best}, P.~N., {et~al.} 2023, \mnras, 523,
  6082

\bibitem[{{Comerford} {et~al.}(2020){Comerford}, {Negus},
  {M{\"u}ller-S{\'a}nchez}, {Eracleous}, {Wylezalek}, {Storchi-Bergmann},
  {Greene}, {Barrows}, {Nevin}, {Roy}, \& {Stemo}}]{2020ApJ...901..159C}
{Comerford}, J.~M., {Negus}, J., {M{\"u}ller-S{\'a}nchez}, F., {et~al.} 2020,
  \apj, 901, 159

\bibitem[{{Condon} {et~al.}(1991){Condon}, {Anderson}, \&
  {Helou}}]{1991ApJ...376...95C}
{Condon}, J.~J., {Anderson}, M.~L., \& {Helou}, G. 1991, \apj, 376, 95

\bibitem[{{Condon} {et~al.}(2002){Condon}, {Cotton}, \&
  {Broderick}}]{2002AJ....124..675C}
{Condon}, J.~J., {Cotton}, W.~D., \& {Broderick}, J.~J. 2002, \aj, 124, 675

\bibitem[{{Condon} {et~al.}(1998){Condon}, {Cotton}, {Greisen}, {Yin},
  {Perley}, {Taylor}, \& {Broderick}}]{1998AJ....115.1693C}
{Condon}, J.~J., {Cotton}, W.~D., {Greisen}, E.~W., {et~al.} 1998, \aj, 115,
  1693

\bibitem[{{Croston} {et~al.}(2019){Croston}, {Hardcastle}, {Mingo}, {Best},
  {Sabater}, {Shimwell}, {Williams}, {Duncan}, {R{\"o}ttgering}, {Brienza},
  {G{\"u}rkan}, {Ineson}, {Miley}, {Morabito}, {O'Sullivan}, \&
  {Prandoni}}]{2019A&A...622A..10C}
{Croston}, J.~H., {Hardcastle}, M.~J., {Mingo}, B., {et~al.} 2019, \aap, 622,
  A10

\bibitem[{{Croton} {et~al.}(2006){Croton}, {Springel}, {White}, {De Lucia},
  {Frenk}, {Gao}, {Jenkins}, {Kauffmann}, {Navarro}, \&
  {Yoshida}}]{2006MNRAS.365...11C}
{Croton}, D.~J., {Springel}, V., {White}, S. D.~M., {et~al.} 2006, \mnras, 365,
  11

\bibitem[{{D'Agostino} {et~al.}(2019){D'Agostino}, {Kewley}, {Groves},
  {Medling}, {Di Teodoro}, {Dopita}, {Thomas}, {Sutherland}, \&
  {Garcia-Burillo}}]{2019MNRAS.487.4153D}
{D'Agostino}, J.~J., {Kewley}, L.~J., {Groves}, B.~A., {et~al.} 2019, \mnras,
  487, 4153

\bibitem[{{Dai} {et~al.}(2018){Dai}, {Wilkes}, {Bergeron}, {Kuraszkiewicz},
  {Omont}, {Atanas}, \& {Teplitz}}]{2018MNRAS.478.4238D}
{Dai}, Y.~S., {Wilkes}, B.~J., {Bergeron}, J., {et~al.} 2018, \mnras, 478, 4238

\bibitem[{{Das} {et~al.}(2024){Das}, {Smith}, {Haskell}, {Hardcastle}, {Best},
  {Duncan}, {Arnaudova}, {Shenoy}, {Kondapally}, {Cochrane}, {Drake},
  {G{\"u}rkan}, {Ma{\l}ek}, {Morabito}, \& {Prandoni}}]{2024MNRAS.531..977D}
{Das}, S., {Smith}, D. J.~B., {Haskell}, P., {et~al.} 2024, \mnras, 531, 977

\bibitem[{{de Jong} {et~al.}(1985){de Jong}, {Klein}, {Wielebinski}, \&
  {Wunderlich}}]{1985A&A...147L...6D}
{de Jong}, T., {Klein}, U., {Wielebinski}, R., \& {Wunderlich}, E. 1985, \aap,
  147, L6

\bibitem[{{Delvecchio} {et~al.}(2021){Delvecchio}, {Daddi}, {Sargent},
  {Jarvis}, {Elbaz}, {Jin}, {Liu}, {Whittam}, {Algera}, {Carraro}, {D'Eugenio},
  {Delhaize}, {Kalita}, {Leslie}, {Moln{\'a}r}, {Novak}, {Prandoni},
  {Smol{\v{c}}i{\'c}}, {Ao}, {Aravena}, {Bournaud}, {Collier},
  {Randriamampandry}, {Randriamanakoto}, {Rodighiero}, {Schober}, {White}, \&
  {Zamorani}}]{2021A&A...647A.123D}
{Delvecchio}, I., {Daddi}, E., {Sargent}, M.~T., {et~al.} 2021, \aap, 647, A123

\bibitem[{{Dey} {et~al.}(2019){Dey}, {Schlegel}, {Lang}, {Blum}, {Burleigh},
  {Fan}, {Findlay}, {Finkbeiner}, {Herrera}, {Juneau}, {Landriau}, {Levi},
  {McGreer}, {Meisner}, {Myers}, {Moustakas}, {Nugent}, {Patej}, {Schlafly},
  {Walker}, {Valdes}, {Weaver}, {Y{\`e}che}, {Zou}, {Zhou}, {Abareshi},
  {Abbott}, {Abolfathi}, {Aguilera}, {Alam}, {Allen}, {Alvarez}, {Annis},
  {Ansarinejad}, {Aubert}, {Beechert}, {Bell}, {BenZvi}, {Beutler}, {Bielby},
  {Bolton}, {Brice{\~n}o}, {Buckley-Geer}, {Butler}, {Calamida}, {Carlberg},
  {Carter}, {Casas}, {Castander}, {Choi}, {Comparat}, {Cukanovaite}, {Delubac},
  {DeVries}, {Dey}, {Dhungana}, {Dickinson}, {Ding}, {Donaldson}, {Duan},
  {Duckworth}, {Eftekharzadeh}, {Eisenstein}, {Etourneau}, {Fagrelius},
  {Farihi}, {Fitzpatrick}, {Font-Ribera}, {Fulmer}, {G{\"a}nsicke},
  {Gaztanaga}, {George}, {Gerdes}, {Gontcho}, {Gorgoni}, {Green}, {Guy},
  {Harmer}, {Hernandez}, {Honscheid}, {Huang}, {James}, {Jannuzi}, {Jiang},
  {Joyce}, {Karcher}, {Karkar}, {Kehoe}, {Kneib}, {Kueter-Young}, {Lan},
  {Lauer}, {Le Guillou}, {Le Van Suu}, {Lee}, {Lesser}, {Perreault Levasseur},
  {Li}, {Mann}, {Marshall}, {Mart{\'\i}nez-V{\'a}zquez}, {Martini}, {du Mas des
  Bourboux}, {McManus}, {Meier}, {M{\'e}nard}, {Metcalfe},
  {Mu{\~n}oz-Guti{\'e}rrez}, {Najita}, {Napier}, {Narayan}, {Newman}, {Nie},
  {Nord}, {Norman}, {Olsen}, {Paat}, {Palanque-Delabrouille}, {Peng},
  {Poppett}, {Poremba}, {Prakash}, {Rabinowitz}, {Raichoor}, {Rezaie},
  {Robertson}, {Roe}, {Ross}, {Ross}, {Rudnick}, {Safonova}, {Saha},
  {S{\'a}nchez}, {Savary}, {Schweiker}, {Scott}, {Seo}, {Shan}, {Silva},
  {Slepian}, {Soto}, {Sprayberry}, {Staten}, {Stillman}, {Stupak}, {Summers},
  {Sien Tie}, {Tirado}, {Vargas-Maga{\~n}a}, {Vivas}, {Wechsler}, {Williams},
  {Yang}, {Yang}, {Yapici}, {Zaritsky}, {Zenteno}, {Zhang}, {Zhang}, {Zhou}, \&
  {Zhou}}]{2019AJ....157..168D}
{Dey}, A., {Schlegel}, D.~J., {Lang}, D., {et~al.} 2019, \aj, 157, 168

\bibitem[{{Di Francesco} {et~al.}(2019){Di Francesco}, {Chalmers}, {Denman},
  {Fissel}, {Friesen}, {Gaensler}, {Hlavacek-Larrondo}, {Kirk}, {Matthews},
  {O'Dea}, {Robishaw}, {Rosolowsky}, {Rupen}, {Sadavoy}, {Sa-Harb}, {Sivakoff},
  {Tahani}, {van der Marel}, {White}, \& {Wilson}}]{2019clrp.2020...32D}
{Di Francesco}, J., {Chalmers}, D., {Denman}, N., {et~al.} 2019, in Canadian
  Long Range Plan for Astronomy and Astrophysics White Papers, Vol. 2020, 32

\bibitem[{{Di Matteo} {et~al.}(2005){Di Matteo}, {Springel}, \&
  {Hernquist}}]{2005Natur.433..604D}
{Di Matteo}, T., {Springel}, V., \& {Hernquist}, L. 2005, \nat, 433, 604

\bibitem[{{Dubois} {et~al.}(2010){Dubois}, {Devriendt}, {Slyz}, \&
  {Teyssier}}]{2010MNRAS.409..985D}
{Dubois}, Y., {Devriendt}, J., {Slyz}, A., \& {Teyssier}, R. 2010, \mnras, 409,
  985

\bibitem[{{Fabian}(2012)}]{2012ARA&A..50..455F}
{Fabian}, A.~C. 2012, \araa, 50, 455

\bibitem[{{Falle}(1991)}]{1991MNRAS.250..581F}
{Falle}, S.~A.~E.~G. 1991, \mnras, 250, 581

\bibitem[{{Fanaroff} \& {Riley}(1974)}]{1974MNRAS.167P..31F}
{Fanaroff}, B.~L. \& {Riley}, J.~M. 1974, \mnras, 167, 31P

\bibitem[{{Ferrarese} \& {Merritt}(2000)}]{2000ApJ...539L...9F}
{Ferrarese}, L. \& {Merritt}, D. 2000, \apjl, 539, L9

\bibitem[{{Foreman-Mackey} {et~al.}(2013){Foreman-Mackey}, {Hogg}, {Lang}, \&
  {Goodman}}]{2013PASP..125..306F}
{Foreman-Mackey}, D., {Hogg}, D.~W., {Lang}, D., \& {Goodman}, J. 2013, \pasp,
  125, 306

\bibitem[{{Gallazzi} {et~al.}(2005){Gallazzi}, {Charlot}, {Brinchmann},
  {White}, \& {Tremonti}}]{2005MNRAS.362...41G}
{Gallazzi}, A., {Charlot}, S., {Brinchmann}, J., {White}, S. D.~M., \&
  {Tremonti}, C.~A. 2005, \mnras, 362, 41

\bibitem[{{Garofalo} \& {Singh}(2019)}]{2019ApJ...871..259G}
{Garofalo}, D. \& {Singh}, C.~B. 2019, \apj, 871, 259

\bibitem[{{Goddard} {et~al.}(2017){Goddard}, {Thomas}, {Maraston}, {Westfall},
  {Etherington}, {Riffel}, {Mallmann}, {Zheng}, {Argudo-Fern{\'a}ndez}, {Lian},
  {Bershady}, {Bundy}, {Drory}, {Law}, {Yan}, {Wake}, {Weijmans}, {Bizyaev},
  {Brownstein}, {Lane}, {Maiolino}, {Masters}, {Merrifield}, {Nitschelm},
  {Pan}, {Roman-Lopes}, {Storchi-Bergmann}, \&
  {Schneider}}]{2017MNRAS.466.4731G}
{Goddard}, D., {Thomas}, D., {Maraston}, C., {et~al.} 2017, \mnras, 466, 4731

\bibitem[{{Guo} {et~al.}(2015){Guo}, {Zheng}, {Wang}, \&
  {Fu}}]{2015ApJ...808L..49G}
{Guo}, K., {Zheng}, X.~Z., {Wang}, T., \& {Fu}, H. 2015, \apjl, 808, L49

\bibitem[{{G{\"u}rkan} {et~al.}(2018){G{\"u}rkan}, {Hardcastle}, {Smith},
  {Best}, {Bourne}, {Calistro-Rivera}, {Heald}, {Jarvis}, {Prandoni},
  {R{\"o}ttgering}, {Sabater}, {Shimwell}, {Tasse}, \&
  {Williams}}]{2018MNRAS.475.3010G}
{G{\"u}rkan}, G., {Hardcastle}, M.~J., {Smith}, D.~J.~B., {et~al.} 2018,
  \mnras, 475, 3010

\bibitem[{{Hale} {et~al.}(2024){Hale}, {Schwarz}, {Best}, {Nakoneczny},
  {Alonso}, {Bacon}, {B{\"o}hme}, {Bhardwaj}, {Bilicki}, {Camera}, {Heneka},
  {Pashapour-Ahmadabadi}, {Tiwari}, {Zheng}, {Duncan}, {Jarvis}, {Kondapally},
  {Magliocchetti}, {Rottgering}, \& {Shimwell}}]{2024MNRAS.527.6540H}
{Hale}, C.~L., {Schwarz}, D.~J., {Best}, P.~N., {et~al.} 2024, \mnras, 527,
  6540

\bibitem[{{Hardcastle} {et~al.}(2023){Hardcastle}, {Horton}, {Williams},
  {Duncan}, {Alegre}, {Barkus}, {Croston}, {Dickinson}, {Osinga},
  {R{\"o}ttgering}, {Sabater}, {Shimwell}, {Smith}, {Best}, {Botteon},
  {Br{\"u}ggen}, {Drabent}, {de Gasperin}, {G{\"u}rkan}, {Hajduk}, {Hale},
  {Hoeft}, {Jamrozy}, {Kunert-Bajraszewska}, {Kondapally}, {Magliocchetti},
  {Mahatma}, {Mostert}, {O'Sullivan}, {Pajdosz-{\'S}mierciak}, {Petley},
  {Pierce}, {Prandoni}, {Schwarz}, {Shulewski}, {Siewert}, {Stott}, {Tang},
  {Vaccari}, {Zheng}, {Bailey}, {Desbled}, {Goyal}, {Gonano}, {Hanset},
  {Kurtz}, {Lim}, {Mielle}, {Molloy}, {Roth}, {Terentev}, \&
  {Torres}}]{2023A&A...678A.151H}
{Hardcastle}, M.~J., {Horton}, M.~A., {Williams}, W.~L., {et~al.} 2023, \aap,
  678, A151

\bibitem[{{Harrison} \& {Ramos Almeida}(2024)}]{2024Galax..12...17H}
{Harrison}, C.~M. \& {Ramos Almeida}, C. 2024, Galaxies, 12, 17

\bibitem[{{Harwood} {et~al.}(2015){Harwood}, {Hardcastle}, \&
  {Croston}}]{2015MNRAS.454.3403H}
{Harwood}, J.~J., {Hardcastle}, M.~J., \& {Croston}, J.~H. 2015, \mnras, 454,
  3403

\bibitem[{{Harwood} {et~al.}(2013){Harwood}, {Hardcastle}, {Croston}, \&
  {Goodger}}]{2013MNRAS.435.3353H}
{Harwood}, J.~J., {Hardcastle}, M.~J., {Croston}, J.~H., \& {Goodger}, J.~L.
  2013, \mnras, 435, 3353

\bibitem[{{Heckman} \& {Best}(2014)}]{2014ARA&A..52..589H}
{Heckman}, T.~M. \& {Best}, P.~N. 2014, \araa, 52, 589

\bibitem[{{Heckman} {et~al.}(2004){Heckman}, {Kauffmann}, {Brinchmann},
  {Charlot}, {Tremonti}, \& {White}}]{2004ApJ...613..109H}
{Heckman}, T.~M., {Kauffmann}, G., {Brinchmann}, J., {et~al.} 2004, \apj, 613,
  109

\bibitem[{{Heesen} {et~al.}(2022){Heesen}, {Staffehl}, {Basu}, {Beck}, {Stein},
  {Tabatabaei}, {Hardcastle}, {Chy{\.z}y}, {Shimwell}, {Adebahr}, {Beswick},
  {Bomans}, {Botteon}, {Brinks}, {Br{\"u}ggen}, {Dettmar}, {Drabent}, {de
  Gasperin}, {G{\"u}rkan}, {Heald}, {Horellou}, {Nikiel-Wroczynski},
  {Paladino}, {Piotrowska}, {R{\"o}ttgering}, {Smith}, \&
  {Tasse}}]{2022A&A...664A..83H}
{Heesen}, V., {Staffehl}, M., {Basu}, A., {et~al.} 2022, \aap, 664, A83

\bibitem[{{Hogg} {et~al.}(2010){Hogg}, {Bovy}, \& {Lang}}]{2010arXiv1008.4686H}
{Hogg}, D.~W., {Bovy}, J., \& {Lang}, D. 2010, arXiv e-prints, arXiv:1008.4686

\bibitem[{{Hotan} {et~al.}(2021){Hotan}, {Bunton}, {Chippendale}, {Whiting},
  {Tuthill}, {Moss}, {McConnell}, {Amy}, {Huynh}, {Allison}, {Anderson},
  {Bannister}, {Bastholm}, {Beresford}, {Bock}, {Bolton}, {Chapman}, {Chow},
  {Collier}, {Cooray}, {Cornwell}, {Diamond}, {Edwards}, {Feain}, {Franzen},
  {George}, {Gupta}, {Hampson}, {Harvey-Smith}, {Hayman}, {Heywood}, {Jacka},
  {Jackson}, {Jackson}, {Jeganathan}, {Johnston}, {Kesteven}, {Kleiner},
  {Koribalski}, {Lee-Waddell}, {Lenc}, {Lensson}, {Mackay}, {Mahony},
  {McClure-Griffiths}, {McConigley}, {Mirtschin}, {Ng}, {Norris}, {Pearce},
  {Phillips}, {Pilawa}, {Raja}, {Reynolds}, {Roberts}, {Roxby}, {Sadler},
  {Shields}, {Schinckel}, {Serra}, {Shaw}, {Sweetnam}, {Troup}, {Tzioumis},
  {Voronkov}, \& {Westmeier}}]{2021PASA...38....9H}
{Hotan}, A.~W., {Bunton}, J.~D., {Chippendale}, A.~P., {et~al.} 2021, \pasa,
  38, e009

\bibitem[{{Ji} \& {Yan}(2020)}]{2020MNRAS.499.5749J}
{Ji}, X. \& {Yan}, R. 2020, \mnras, 499, 5749

\bibitem[{{Jin} {et~al.}(2021){Jin}, {Dai}, {Pan}, {Lin}, {Li}, {Hsieh},
  {Shen}, {Yuan}, {Feng}, {Cheng}, {Xu}, {Huang}, \&
  {Zhang}}]{2021ApJ...923....6J}
{Jin}, G., {Dai}, Y.~S., {Pan}, H.-A., {et~al.} 2021, \apj, 923, 6

\bibitem[{{Jonas} \& {MeerKAT Team}(2016)}]{2016mks..confE...1J}
{Jonas}, J. \& {MeerKAT Team}. 2016, in MeerKAT Science: On the Pathway to the
  SKA, 1

\bibitem[{{Kauffmann} \& {Haehnelt}(2000)}]{2000MNRAS.311..576K}
{Kauffmann}, G. \& {Haehnelt}, M. 2000, \mnras, 311, 576

\bibitem[{{Kauffmann} {et~al.}(2003){Kauffmann}, {Heckman}, {Tremonti},
  {Brinchmann}, {Charlot}, {White}, {Ridgway}, {Brinkmann}, {Fukugita}, {Hall},
  {Ivezi{\'c}}, {Richards}, \& {Schneider}}]{2003MNRAS.346.1055K}
{Kauffmann}, G., {Heckman}, T.~M., {Tremonti}, C., {et~al.} 2003, \mnras, 346,
  1055

\bibitem[{{Kennicutt} \& {Evans}(2012)}]{2012ARA&A..50..531K}
{Kennicutt}, R.~C. \& {Evans}, N.~J. 2012, \araa, 50, 531

\bibitem[{{Klindt} {et~al.}(2019){Klindt}, {Alexander}, {Rosario}, {Lusso}, \&
  {Fotopoulou}}]{2019MNRAS.488.3109K}
{Klindt}, L., {Alexander}, D.~M., {Rosario}, D.~J., {Lusso}, E., \&
  {Fotopoulou}, S. 2019, \mnras, 488, 3109

\bibitem[{{Li} {et~al.}(2024){Li}, {Dai}, {Huang}, {Wuyts}, \&
  {Cao}}]{2024ApJ...963...99L}
{Li}, Z.-J., {Dai}, Y.~S., {Huang}, J.~S., {Wuyts}, S., \& {Cao}, T.-W. 2024,
  \apj, 963, 99

\bibitem[{{Madau} \& {Dickinson}(2014)}]{2014ARA&A..52..415M}
{Madau}, P. \& {Dickinson}, M. 2014, \araa, 52, 415

\bibitem[{{Magliocchetti}(2022)}]{2022A&ARv..30....6M}
{Magliocchetti}, M. 2022, \aapr, 30, 6

\bibitem[{{Magorrian} {et~al.}(1998){Magorrian}, {Tremaine}, {Richstone},
  {Bender}, {Bower}, {Dressler}, {Faber}, {Gebhardt}, {Green}, {Grillmair},
  {Kormendy}, \& {Lauer}}]{1998AJ....115.2285M}
{Magorrian}, J., {Tremaine}, S., {Richstone}, D., {et~al.} 1998, \aj, 115, 2285

\bibitem[{{Ma{\l}ek} {et~al.}(2018){Ma{\l}ek}, {Buat}, {Roehlly}, {Burgarella},
  {Hurley}, {Shirley}, {Duncan}, {Efstathiou}, {Papadopoulos}, {Vaccari},
  {Farrah}, {Marchetti}, \& {Oliver}}]{2018A&A...620A..50M}
{Ma{\l}ek}, K., {Buat}, V., {Roehlly}, Y., {et~al.} 2018, \aap, 620, A50

\bibitem[{{Maraston} {et~al.}(2020){Maraston}, {Hill}, {Thomas}, {Yan}, {Chen},
  {Lian}, {Parikh}, {Neumann}, {Meneses-Goytia}, {Bershady}, {Drory},
  {Bizyaev}, {Concas}, {Brownstein}, {Lazarz}, {Stringfellow}, \&
  {Stassun}}]{2020MNRAS.496.2962M}
{Maraston}, C., {Hill}, L., {Thomas}, D., {et~al.} 2020, \mnras, 496, 2962

\bibitem[{{McConnell} \& {Ma}(2013)}]{2013ApJ...764..184M}
{McConnell}, N.~J. \& {Ma}, C.-P. 2013, \apj, 764, 184

\bibitem[{{Mohan} \& {Rafferty}(2015)}]{2015ascl.soft02007M}
{Mohan}, N. \& {Rafferty}, D. 2015, {PyBDSF: Python Blob Detection and Source
  Finder}, Astrophysics Source Code Library, record ascl:1502.007

\bibitem[{{Mulcahey} {et~al.}(2022){Mulcahey}, {Leslie}, {Jackson}, {Young},
  {Prandoni}, {Hardcastle}, {Roy}, {Ma{\l}ek}, {Magliocchetti}, {Bonato},
  {R{\"o}ttgering}, \& {Drabent}}]{2022A&A...665A.144M}
{Mulcahey}, C.~R., {Leslie}, S.~K., {Jackson}, T.~M., {et~al.} 2022, \aap, 665,
  A144

\bibitem[{{Nelson} {et~al.}(2019){Nelson}, {Pillepich}, {Springel}, {Pakmor},
  {Weinberger}, {Genel}, {Torrey}, {Vogelsberger}, {Marinacci}, \&
  {Hernquist}}]{2019MNRAS.490.3234N}
{Nelson}, D., {Pillepich}, A., {Springel}, V., {et~al.} 2019, \mnras, 490, 3234

\bibitem[{{Neumann} {et~al.}(2022){Neumann}, {Thomas}, {Maraston}, {Hill},
  {Nanni}, {Wenman}, {Lian}, {Comparat}, {Gonzalez-Perez}, {Westfall}, {Yan},
  {Chen}, {Stringfellow}, {Bershady}, {Brownstein}, {Drory}, \&
  {Schneider}}]{2022MNRAS.513.5988N}
{Neumann}, J., {Thomas}, D., {Maraston}, C., {et~al.} 2022, \mnras, 513, 5988

\bibitem[{{Osterbrock}(1989)}]{1989agna.book.....O}
{Osterbrock}, D.~E. 1989, {Astrophysics of gaseous nebulae and active galactic
  nuclei}

\bibitem[{{Pacifici} {et~al.}(2016){Pacifici}, {Oh}, {Oh}, {Lee}, \&
  {Yi}}]{2016ApJ...824...45P}
{Pacifici}, C., {Oh}, S., {Oh}, K., {Lee}, J., \& {Yi}, S.~K. 2016, \apj, 824,
  45

\bibitem[{{Padovani} {et~al.}(2017){Padovani}, {Alexander}, {Assef}, {De
  Marco}, {Giommi}, {Hickox}, {Richards}, {Smol{\v{c}}i{\'c}},
  {Hatziminaoglou}, {Mainieri}, \& {Salvato}}]{2017A&ARv..25....2P}
{Padovani}, P., {Alexander}, D.~M., {Assef}, R.~J., {et~al.} 2017, \aapr, 25, 2

\bibitem[{{Peng} {et~al.}(2010){Peng}, {Lilly}, {Kova{\v{c}}}, {Bolzonella},
  {Pozzetti}, {Renzini}, {Zamorani}, {Ilbert}, {Knobel}, {Iovino}, {Maier},
  {Cucciati}, {Tasca}, {Carollo}, {Silverman}, {Kampczyk}, {de Ravel},
  {Sanders}, {Scoville}, {Contini}, {Mainieri}, {Scodeggio}, {Kneib}, {Le
  F{\`e}vre}, {Bardelli}, {Bongiorno}, {Caputi}, {Coppa}, {de la Torre},
  {Franzetti}, {Garilli}, {Lamareille}, {Le Borgne}, {Le Brun}, {Mignoli},
  {Perez Montero}, {Pello}, {Ricciardelli}, {Tanaka}, {Tresse}, {Vergani},
  {Welikala}, {Zucca}, {Oesch}, {Abbas}, {Barnes}, {Bordoloi}, {Bottini},
  {Cappi}, {Cassata}, {Cimatti}, {Fumana}, {Hasinger}, {Koekemoer},
  {Leauthaud}, {Maccagni}, {Marinoni}, {McCracken}, {Memeo}, {Meneux}, {Nair},
  {Porciani}, {Presotto}, \& {Scaramella}}]{2010ApJ...721..193P}
{Peng}, Y.-j., {Lilly}, S.~J., {Kova{\v{c}}}, K., {et~al.} 2010, \apj, 721, 193

\bibitem[{{Penney} {et~al.}(2020){Penney}, {Blain}, {Assef}, {Diaz-Santos},
  {Gonz{\'a}lez-L{\'o}pez}, {Tsai}, {Aravena}, {Eisenhardt}, {Jones}, {Jun},
  {Kim}, {Stern}, \& {Wu}}]{2020MNRAS.496.1565P}
{Penney}, J.~I., {Blain}, A.~W., {Assef}, R.~J., {et~al.} 2020, \mnras, 496,
  1565

\bibitem[{{Piotrowska} {et~al.}(2022){Piotrowska}, {Bluck}, {Maiolino}, \&
  {Peng}}]{2022MNRAS.512.1052P}
{Piotrowska}, J.~M., {Bluck}, A. F.~L., {Maiolino}, R., \& {Peng}, Y. 2022,
  \mnras, 512, 1052

\bibitem[{{Rodriguez-Gomez} {et~al.}(2016){Rodriguez-Gomez}, {Pillepich},
  {Sales}, {Genel}, {Vogelsberger}, {Zhu}, {Wellons}, {Nelson}, {Torrey},
  {Springel}, {Ma}, \& {Hernquist}}]{2016MNRAS.458.2371R}
{Rodriguez-Gomez}, V., {Pillepich}, A., {Sales}, L.~V., {et~al.} 2016, \mnras,
  458, 2371

\bibitem[{{Rothberg} \& {Joseph}(2004)}]{2004AJ....128.2098R}
{Rothberg}, B. \& {Joseph}, R.~D. 2004, \aj, 128, 2098

\bibitem[{{Russell} {et~al.}(2013){Russell}, {McNamara}, {Edge}, {Hogan},
  {Main}, \& {Vantyghem}}]{2013MNRAS.432..530R}
{Russell}, H.~R., {McNamara}, B.~R., {Edge}, A.~C., {et~al.} 2013, \mnras, 432,
  530

\bibitem[{{Sabater} {et~al.}(2019){Sabater}, {Best}, {Hardcastle}, {Shimwell},
  {Tasse}, {Williams}, {Br{\"u}ggen}, {Cochrane}, {Croston}, {de Gasperin},
  {Duncan}, {G{\"u}rkan}, {Mechev}, {Morabito}, {Prandoni}, {R{\"o}ttgering},
  {Smith}, {Harwood}, {Mingo}, {Mooney}, \& {Saxena}}]{2019A&A...622A..17S}
{Sabater}, J., {Best}, P.~N., {Hardcastle}, M.~J., {et~al.} 2019, \aap, 622,
  A17

\bibitem[{{Saintonge} \& {Catinella}(2022)}]{2022ARA&A..60..319S}
{Saintonge}, A. \& {Catinella}, B. 2022, \araa, 60, 319

\bibitem[{{Salom{\'e}} {et~al.}(2015){Salom{\'e}}, {Salom{\'e}}, \&
  {Combes}}]{2015A&A...574A..34S}
{Salom{\'e}}, Q., {Salom{\'e}}, P., \& {Combes}, F. 2015, \aap, 574, A34

\bibitem[{{S{\'a}nchez} {et~al.}(2022){S{\'a}nchez}, {Barrera-Ballesteros},
  {Lacerda}, {Mej{\'\i}a-Narvaez}, {Camps-Fari{\~n}a}, {Bruzual},
  {Espinosa-Ponce}, {Rodr{\'\i}guez-Puebla}, {Calette}, {Ibarra-Medel},
  {Avila-Reese}, {Hernandez-Toledo}, {Bershady}, {Cano-Diaz}, \&
  {Munguia-Cordova}}]{2022ApJS..262...36S}
{S{\'a}nchez}, S.~F., {Barrera-Ballesteros}, J.~K., {Lacerda}, E., {et~al.}
  2022, \apjs, 262, 36

\bibitem[{{Schaye} {et~al.}(2015){Schaye}, {Crain}, {Bower}, {Furlong},
  {Schaller}, {Theuns}, {Dalla Vecchia}, {Frenk}, {McCarthy}, {Helly},
  {Jenkins}, {Rosas-Guevara}, {White}, {Baes}, {Booth}, {Camps}, {Navarro},
  {Qu}, {Rahmati}, {Sawala}, {Thomas}, \& {Trayford}}]{2015MNRAS.446..521S}
{Schaye}, J., {Crain}, R.~A., {Bower}, R.~G., {et~al.} 2015, \mnras, 446, 521

\bibitem[{{Schlafly} {et~al.}(2019){Schlafly}, {Meisner}, \&
  {Green}}]{2019ApJS..240...30S}
{Schlafly}, E.~F., {Meisner}, A.~M., \& {Green}, G.~M. 2019, \apjs, 240, 30

\bibitem[{{Schutte} \& {Reines}(2022)}]{2022Natur.601..329S}
{Schutte}, Z. \& {Reines}, A.~E. 2022, \nat, 601, 329

\bibitem[{{Shimwell} {et~al.}(2022){Shimwell}, {Hardcastle}, {Tasse}, {Best},
  {R{\"o}ttgering}, {Williams}, {Botteon}, {Drabent}, {Mechev}, {Shulevski},
  {van Weeren}, {Bester}, {Br{\"u}ggen}, {Brunetti}, {Callingham}, {Chy{\.z}y},
  {Conway}, {Dijkema}, {Duncan}, {de Gasperin}, {Hale}, {Haverkorn}, {Hugo},
  {Jackson}, {Mevius}, {Miley}, {Morabito}, {Morganti}, {Offringa}, {Oonk},
  {Rafferty}, {Sabater}, {Smith}, {Schwarz}, {Smirnov}, {O'Sullivan},
  {Vedantham}, {White}, {Albert}, {Alegre}, {Asabere}, {Bacon}, {Bonafede},
  {Bonnassieux}, {Brienza}, {Bilicki}, {Bonato}, {Calistro Rivera}, {Cassano},
  {Cochrane}, {Croston}, {Cuciti}, {Dallacasa}, {Danezi}, {Dettmar}, {Di
  Gennaro}, {Edler}, {En{\ss}lin}, {Emig}, {Franzen}, {Garc{\'\i}a-Vergara},
  {Grange}, {G{\"u}rkan}, {Hajduk}, {Heald}, {Heesen}, {Hoang}, {Hoeft},
  {Horellou}, {Iacobelli}, {Jamrozy}, {Jeli{\'c}}, {Kondapally}, {Kukreti},
  {Kunert-Bajraszewska}, {Magliocchetti}, {Mahatma}, {Ma{\l}ek}, {Mandal},
  {Massaro}, {Meyer-Zhao}, {Mingo}, {Mostert}, {Nair}, {Nakoneczny},
  {Nikiel-Wroczy{\'n}ski}, {Orr{\'u}}, {Pajdosz-{\'S}mierciak}, {Pasini},
  {Prandoni}, {van Piggelen}, {Rajpurohit}, {Retana-Montenegro}, {Riseley},
  {Rowlinson}, {Saxena}, {Schrijvers}, {Sweijen}, {Siewert}, {Timmerman},
  {Vaccari}, {Vink}, {West}, {Wo{\l}owska}, {Zhang}, \&
  {Zheng}}]{2022A&A...659A...1S}
{Shimwell}, T.~W., {Hardcastle}, M.~J., {Tasse}, C., {et~al.} 2022, \aap, 659,
  A1

\bibitem[{{Shimwell} {et~al.}(2017){Shimwell}, {R{\"o}ttgering}, {Best},
  {Williams}, {Dijkema}, {de Gasperin}, {Hardcastle}, {Heald}, {Hoang},
  {Horneffer}, {Intema}, {Mahony}, {Mandal}, {Mechev}, {Morabito}, {Oonk},
  {Rafferty}, {Retana-Montenegro}, {Sabater}, {Tasse}, {van Weeren},
  {Br{\"u}ggen}, {Brunetti}, {Chy{\.z}y}, {Conway}, {Haverkorn}, {Jackson},
  {Jarvis}, {McKean}, {Miley}, {Morganti}, {White}, {Wise}, {van Bemmel},
  {Beck}, {Brienza}, {Bonafede}, {Calistro Rivera}, {Cassano}, {Clarke},
  {Cseh}, {Deller}, {Drabent}, {van Driel}, {Engels}, {Falcke}, {Ferrari},
  {Fr{\"o}hlich}, {Garrett}, {Harwood}, {Heesen}, {Hoeft}, {Horellou},
  {Israel}, {Kapi{\'n}ska}, {Kunert-Bajraszewska}, {McKay}, {Mohan},
  {Orr{\'u}}, {Pizzo}, {Prandoni}, {Schwarz}, {Shulevski}, {Sipior}, {Smith},
  {Sridhar}, {Steinmetz}, {Stroe}, {Varenius}, {van der Werf}, {Zensus}, \&
  {Zwart}}]{2017A&A...598A.104S}
{Shimwell}, T.~W., {R{\"o}ttgering}, H.~J.~A., {Best}, P.~N., {et~al.} 2017,
  \aap, 598, A104

\bibitem[{{Silk} {et~al.}(2024){Silk}, {Begelman}, {Norman}, {Nusser}, \&
  {Wyse}}]{2024ApJ...961L..39S}
{Silk}, J., {Begelman}, M.~C., {Norman}, C., {Nusser}, A., \& {Wyse}, R. F.~G.
  2024, \apjl, 961, L39

\bibitem[{{Silverman} {et~al.}(2008){Silverman}, {Green}, {Barkhouse}, {Kim},
  {Kim}, {Wilkes}, {Cameron}, {Hasinger}, {Jannuzi}, {Smith}, {Smith}, \&
  {Tananbaum}}]{2008ApJ...679..118S}
{Silverman}, J.~D., {Green}, P.~J., {Barkhouse}, W.~A., {et~al.} 2008, \apj,
  679, 118

\bibitem[{{Smee} {et~al.}(2013){Smee}, {Gunn}, {Uomoto}, {Roe}, {Schlegel},
  {Rockosi}, {Carr}, {Leger}, {Dawson}, {Olmstead}, {Brinkmann}, {Owen},
  {Barkhouser}, {Honscheid}, {Harding}, {Long}, {Lupton}, {Loomis}, {Anderson},
  {Annis}, {Bernardi}, {Bhardwaj}, {Bizyaev}, {Bolton}, {Brewington}, {Briggs},
  {Burles}, {Burns}, {Castander}, {Connolly}, {Davenport}, {Ebelke}, {Epps},
  {Feldman}, {Friedman}, {Frieman}, {Heckman}, {Hull}, {Knapp}, {Lawrence},
  {Loveday}, {Mannery}, {Malanushenko}, {Malanushenko}, {Merrelli}, {Muna},
  {Newman}, {Nichol}, {Oravetz}, {Pan}, {Pope}, {Ricketts}, {Shelden},
  {Sandford}, {Siegmund}, {Simmons}, {Smith}, {Snedden}, {Schneider},
  {SubbaRao}, {Tremonti}, {Waddell}, \& {York}}]{2013AJ....146...32S}
{Smee}, S.~A., {Gunn}, J.~E., {Uomoto}, A., {et~al.} 2013, \aj, 146, 32

\bibitem[{{Smith} {et~al.}(2021){Smith}, {Haskell}, {G{\"u}rkan}, {Best},
  {Hardcastle}, {Kondapally}, {Williams}, {Duncan}, {Cochrane}, {McCheyne},
  {R{\"o}ttgering}, {Sabater}, {Shimwell}, {Tasse}, {Bonato}, {Bondi},
  {Jarvis}, {Leslie}, {Prandoni}, \& {Wang}}]{2021A&A...648A...6S}
{Smith}, D.~J.~B., {Haskell}, P., {G{\"u}rkan}, G., {et~al.} 2021, \aap, 648,
  A6

\bibitem[{{Springel} {et~al.}(2005){Springel}, {Di Matteo}, \&
  {Hernquist}}]{2005MNRAS.361..776S}
{Springel}, V., {Di Matteo}, T., \& {Hernquist}, L. 2005, \mnras, 361, 776

\bibitem[{{Steffen} {et~al.}(2023){Steffen}, {Fu}, {Brownstein}, {Comerford},
  {Cruz-Gonz{\'a}lez}, {Sophia Dai}, {Drory}, {Gross}, {Alenka Negrete}, \&
  {Yan}}]{2023ApJ...942..107S}
{Steffen}, J.~L., {Fu}, H., {Brownstein}, J.~R., {et~al.} 2023, \apj, 942, 107

\bibitem[{{Sun} \& {Malkan}(1989)}]{1989ApJ...346...68S}
{Sun}, W.-H. \& {Malkan}, M.~A. 1989, \apj, 346, 68

\bibitem[{{Torrey} {et~al.}(2014){Torrey}, {Vogelsberger}, {Genel}, {Sijacki},
  {Springel}, \& {Hernquist}}]{2014MNRAS.438.1985T}
{Torrey}, P., {Vogelsberger}, M., {Genel}, S., {et~al.} 2014, \mnras, 438, 1985

\bibitem[{{Turner} {et~al.}(2018){Turner}, {Shabala}, \&
  {Krause}}]{2018MNRAS.474.3361T}
{Turner}, R.~J., {Shabala}, S.~S., \& {Krause}, M. G.~H. 2018, \mnras, 474,
  3361

\bibitem[{{van Haarlem} {et~al.}(2013){van Haarlem}, {Wise}, {Gunst}, {Heald},
  {McKean}, {Hessels}, {de Bruyn}, {Nijboer}, {Swinbank}, {Fallows},
  {Brentjens}, {Nelles}, {Beck}, {Falcke}, {Fender}, {H{\"o}randel},
  {Koopmans}, {Mann}, {Miley}, {R{\"o}ttgering}, {Stappers}, {Wijers},
  {Zaroubi}, {van den Akker}, {Alexov}, {Anderson}, {Anderson}, {van Ardenne},
  {Arts}, {Asgekar}, {Avruch}, {Batejat}, {B{\"a}hren}, {Bell}, {Bell}, {van
  Bemmel}, {Bennema}, {Bentum}, {Bernardi}, {Best}, {B{\^\i}rzan}, {Bonafede},
  {Boonstra}, {Braun}, {Bregman}, {Breitling}, {van de Brink}, {Broderick},
  {Broekema}, {Brouw}, {Br{\"u}ggen}, {Butcher}, {van Cappellen}, {Ciardi},
  {Coenen}, {Conway}, {Coolen}, {Corstanje}, {Damstra}, {Davies}, {Deller},
  {Dettmar}, {van Diepen}, {Dijkstra}, {Donker}, {Doorduin}, {Dromer}, {Drost},
  {van Duin}, {Eisl{\"o}ffel}, {van Enst}, {Ferrari}, {Frieswijk}, {Gankema},
  {Garrett}, {de Gasperin}, {Gerbers}, {de Geus}, {Grie{\ss}meier}, {Grit},
  {Gruppen}, {Hamaker}, {Hassall}, {Hoeft}, {Holties}, {Horneffer}, {van der
  Horst}, {van Houwelingen}, {Huijgen}, {Iacobelli}, {Intema}, {Jackson},
  {Jelic}, {de Jong}, {Juette}, {Kant}, {Karastergiou}, {Koers}, {Kollen},
  {Kondratiev}, {Kooistra}, {Koopman}, {Koster}, {Kuniyoshi}, {Kramer},
  {Kuper}, {Lambropoulos}, {Law}, {van Leeuwen}, {Lemaitre}, {Loose}, {Maat},
  {Macario}, {Markoff}, {Masters}, {McFadden}, {McKay-Bukowski}, {Meijering},
  {Meulman}, {Mevius}, {Middelberg}, {Millenaar}, {Miller-Jones}, {Mohan},
  {Mol}, {Morawietz}, {Morganti}, {Mulcahy}, {Mulder}, {Munk}, {Nieuwenhuis},
  {van Nieuwpoort}, {Noordam}, {Norden}, {Noutsos}, {Offringa}, {Olofsson},
  {Omar}, {Orr{\'u}}, {Overeem}, {Paas}, {Pandey-Pommier}, {Pandey}, {Pizzo},
  {Polatidis}, {Rafferty}, {Rawlings}, {Reich}, {de Reijer}, {Reitsma},
  {Renting}, {Riemers}, {Rol}, {Romein}, {Roosjen}, {Ruiter}, {Scaife}, {van
  der Schaaf}, {Scheers}, {Schellart}, {Schoenmakers}, {Schoonderbeek},
  {Serylak}, {Shulevski}, {Sluman}, {Smirnov}, {Sobey}, {Spreeuw}, {Steinmetz},
  {Sterks}, {Stiepel}, {Stuurwold}, {Tagger}, {Tang}, {Tasse}, {Thomas},
  {Thoudam}, {Toribio}, {van der Tol}, {Usov}, {van Veelen}, {van der Veen},
  {ter Veen}, {Verbiest}, {Vermeulen}, {Vermaas}, {Vocks}, {Vogt}, {de Vos},
  {van der Wal}, {van Weeren}, {Weggemans}, {Weltevrede}, {White}, {Wijnholds},
  {Wilhelmsson}, {Wucknitz}, {Yatawatta}, {Zarka}, {Zensus}, \& {van
  Zwieten}}]{2013A&A...556A...2V}
{van Haarlem}, M.~P., {Wise}, M.~W., {Gunst}, A.~W., {et~al.} 2013, \aap, 556,
  A2

\bibitem[{{V{\'a}zquez-Mata} {et~al.}(2022){V{\'a}zquez-Mata},
  {Hern{\'a}ndez-Toledo}, {Avila-Reese}, {Herrera-Endoqui},
  {Rodr{\'\i}guez-Puebla}, {Cano-D{\'\i}az}, {Lacerna},
  {Mart{\'\i}nez-V{\'a}zquez}, \& {Lane}}]{2022MNRAS.512.2222V}
{V{\'a}zquez-Mata}, J.~A., {Hern{\'a}ndez-Toledo}, H.~M., {Avila-Reese}, V.,
  {et~al.} 2022, \mnras, 512, 2222

\bibitem[{{Wake} {et~al.}(2017){Wake}, {Bundy}, {Diamond-Stanic}, {Yan},
  {Blanton}, {Bershady}, {S{\'a}nchez-Gallego}, {Drory}, {Jones}, {Kauffmann},
  {Law}, {Li}, {MacDonald}, {Masters}, {Thomas}, {Tinker}, {Weijmans}, \&
  {Brownstein}}]{2017AJ....154...86W}
{Wake}, D.~A., {Bundy}, K., {Diamond-Stanic}, A.~M., {et~al.} 2017, \aj, 154,
  86

\bibitem[{{Wang} {et~al.}(2016){Wang}, {Norberg}, {Gunawardhana}, {Heinis},
  {Baldry}, {Bland-Hawthorn}, {Bourne}, {Brough}, {Brown}, {Cluver}, {Cooray},
  {da Cunha}, {Driver}, {Dunne}, {Dye}, {Eales}, {Grootes}, {Holwerda},
  {Hopkins}, {Ibar}, {Ivison}, {Lacey}, {Lara-Lopez}, {Loveday}, {Maddox},
  {Micha{\l}owski}, {Oteo}, {Owers}, {Popescu}, {Smith}, {Taylor}, {Tuffs}, \&
  {van der Werf}}]{2016MNRAS.461.1898W}
{Wang}, L., {Norberg}, P., {Gunawardhana}, M.~L.~P., {et~al.} 2016, \mnras,
  461, 1898

\bibitem[{{Weinberger} {et~al.}(2017){Weinberger}, {Springel}, {Hernquist},
  {Pillepich}, {Marinacci}, {Pakmor}, {Nelson}, {Genel}, {Vogelsberger},
  {Naiman}, \& {Torrey}}]{2017MNRAS.465.3291W}
{Weinberger}, R., {Springel}, V., {Hernquist}, L., {et~al.} 2017, \mnras, 465,
  3291

\bibitem[{{Westfall} {et~al.}(2019){Westfall}, {Cappellari}, {Bershady},
  {Bundy}, {Belfiore}, {Ji}, {Law}, {Schaefer}, {Shetty}, {Tremonti}, {Yan},
  {Andrews}, {Brownstein}, {Cherinka}, {Coccato}, {Drory}, {Maraston},
  {Parikh}, {S{\'a}nchez-Gallego}, {Thomas}, {Weijmans}, {Barrera-Ballesteros},
  {Du}, {Goddard}, {Li}, {Masters}, {Ibarra Medel}, {S{\'a}nchez}, {Yang},
  {Zheng}, \& {Zhou}}]{2019AJ....158..231W}
{Westfall}, K.~B., {Cappellari}, M., {Bershady}, M.~A., {et~al.} 2019, \aj,
  158, 231

\bibitem[{{Wilkinson} {et~al.}(2017){Wilkinson}, {Maraston}, {Goddard},
  {Thomas}, \& {Parikh}}]{2017MNRAS.472.4297W}
{Wilkinson}, D.~M., {Maraston}, C., {Goddard}, D., {Thomas}, D., \& {Parikh},
  T. 2017, \mnras, 472, 4297

\bibitem[{{Williams} {et~al.}(2019){Williams}, {Hardcastle}, {Best}, {Sabater},
  {Croston}, {Duncan}, {Shimwell}, {R{\"o}ttgering}, {Nisbet}, {G{\"u}rkan},
  {Alegre}, {Cochrane}, {Goyal}, {Hale}, {Jackson}, {Jamrozy}, {Kondapally},
  {Kunert-Bajraszewska}, {Mahatma}, {Mingo}, {Morabito}, {Prandoni},
  {Roskowinski}, {Shulevski}, {Smith}, {Tasse}, {Urquhart}, {Webster}, {White},
  {Beswick}, {Callingham}, {Chy{\.z}y}, {de Gasperin}, {Harwood}, {Hoeft},
  {Iacobelli}, {McKean}, {Mechev}, {Miley}, {Schwarz}, \& {van
  Weeren}}]{2019A&A...622A...2W}
{Williams}, W.~L., {Hardcastle}, M.~J., {Best}, P.~N., {et~al.} 2019, \aap,
  622, A2

\bibitem[{{Yan} {et~al.}(2019){Yan}, {Chen}, {Lazarz}, {Bizyaev}, {Maraston},
  {Stringfellow}, {McCarthy}, {Meneses-Goytia}, {Law}, {Thomas}, {Falcon
  Barroso}, {S{\'a}nchez-Gallego}, {Schlafly}, {Zheng}, {Argudo-Fern{\'a}ndez},
  {Beaton}, {Beers}, {Bershady}, {Blanton}, {Brownstein}, {Bundy}, {Chambers},
  {Cherinka}, {De Lee}, {Drory}, {Galbany}, {Holtzman}, {Imig}, {Kaiser},
  {Kinemuchi}, {Liu}, {Luo}, {Magnier}, {Majewski}, {Nair}, {Oravetz},
  {Oravetz}, {Pan}, {Sobeck}, {Stassun}, {Talbot}, {Tremonti}, {Waters},
  {Weijmans}, {Wilhelm}, {Zasowski}, {Zhao}, \& {Zhao}}]{2019ApJ...883..175Y}
{Yan}, R., {Chen}, Y., {Lazarz}, D., {et~al.} 2019, \apj, 883, 175

\bibitem[{{Yang} \& {Reynolds}(2016)}]{2016ApJ...829...90Y}
{Yang}, H. Y.~K. \& {Reynolds}, C.~S. 2016, \apj, 829, 90

\bibitem[{{Zheng} {et~al.}(2023){Zheng}, {R{\"o}ttgering}, {van der Wel}, \&
  {Cappellari}}]{2023A&A...673A..12Z}
{Zheng}, X., {R{\"o}ttgering}, H., {van der Wel}, A., \& {Cappellari}, M. 2023,
  \aap, 673, A12

\end{thebibliography}
%
\begin{appendix} 
\section{Fitting the $L_{\rm 144MHz}$-SFR relation}
\label{ap:fit}

To build the $L_{\rm 144MHz}$-SFR relation, the key is to select a sample of pure star-forming galaxies with reliable SFR measurement and without radio jet contribution.

Using the integral field spectra from MaNGA, the SFR can be derived from the H$\alpha$ line, and corrected for dust attenuation on sub-galactic scales.
There are galaxies in which the H$\alpha$ is not dominated by star formation but instead is possibly contaminated by narrow-line AGN or diffuse ionized gas.
We exclude these galaxies using the widely used emission line diagnostic \citepads[BPT diagram,][]{1981PASP...93....5B}
based on the [\ion{N}{II}]$\lambda$6585/H$\alpha$
and [\ion{O}{III}]$\lambda$5008/H$\beta$ line ratios measured from the central 2.5\arcsec \ MaNGA spaxels.
Specifically, our `pure-SFG' sample only includes galaxies locating in the star-forming region of the BPT diagram,
using the separation diagnostic derived by \citepads{2003MNRAS.346.1055K}, 
and the galaxies should also have nuclear H$\alpha$ equivalent widths larger than 3$\AA$, which ensures the SFR calculation is not affected by the diffused gas from evolved stars \citepads{2011MNRAS.413.1687C}.

\begin{figure}
\centering
\includegraphics[width=\hsize]{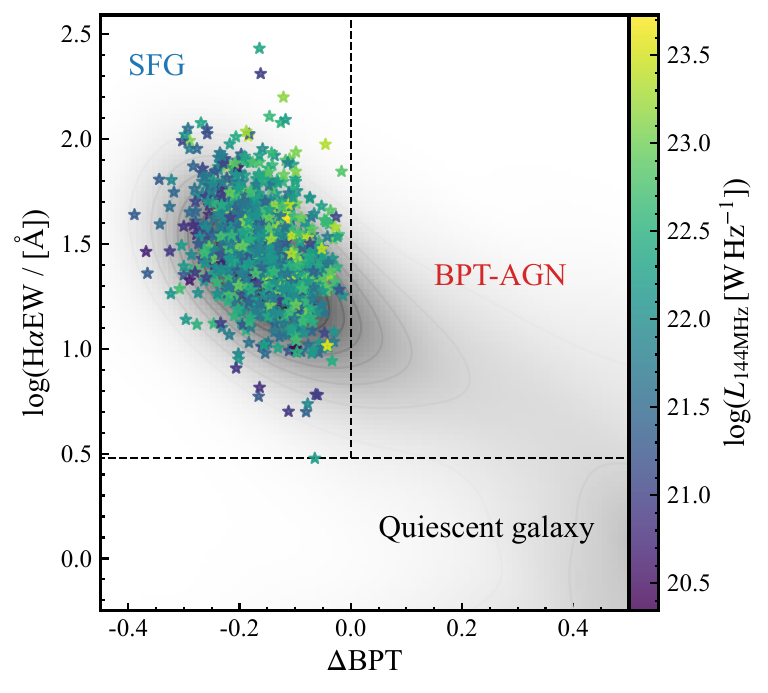}
  \caption{The `pure SFG' sample selected by the emission line diagnostic and radio morphology.
  The y-axis is the equivalent width of H$\alpha$ in the nuclear region measured from MaNGA spectra.
  The x-axis is the distance to the division line in the BPT diagram \citepads[Equation 1 in][]{2003MNRAS.346.1055K},
  based on the [\ion{N}{II}]$\lambda$6585/H$\alpha$
  and [\ion{O}{III}]$\lambda$5008/H$\beta$ ratios.
  Negative and positive $\Delta$BPT mean that galaxies locate in the 
 `star-forming' region and `AGN' region in BPT diagram, respectively.
 This H$\alpha$-$\Delta$BPT diagram can classify the galaxies into SFG, AGN, and quiescent galaxies. Our `pure SFG' sample is plotted as stars and color-coded by their radio luminosities. Background gray-scale contour is the distribution of the whole MaNGA sample.}
     \label{fig:dbpt}
\end{figure} 

\begin{figure}
\centering
\includegraphics[width=\hsize]{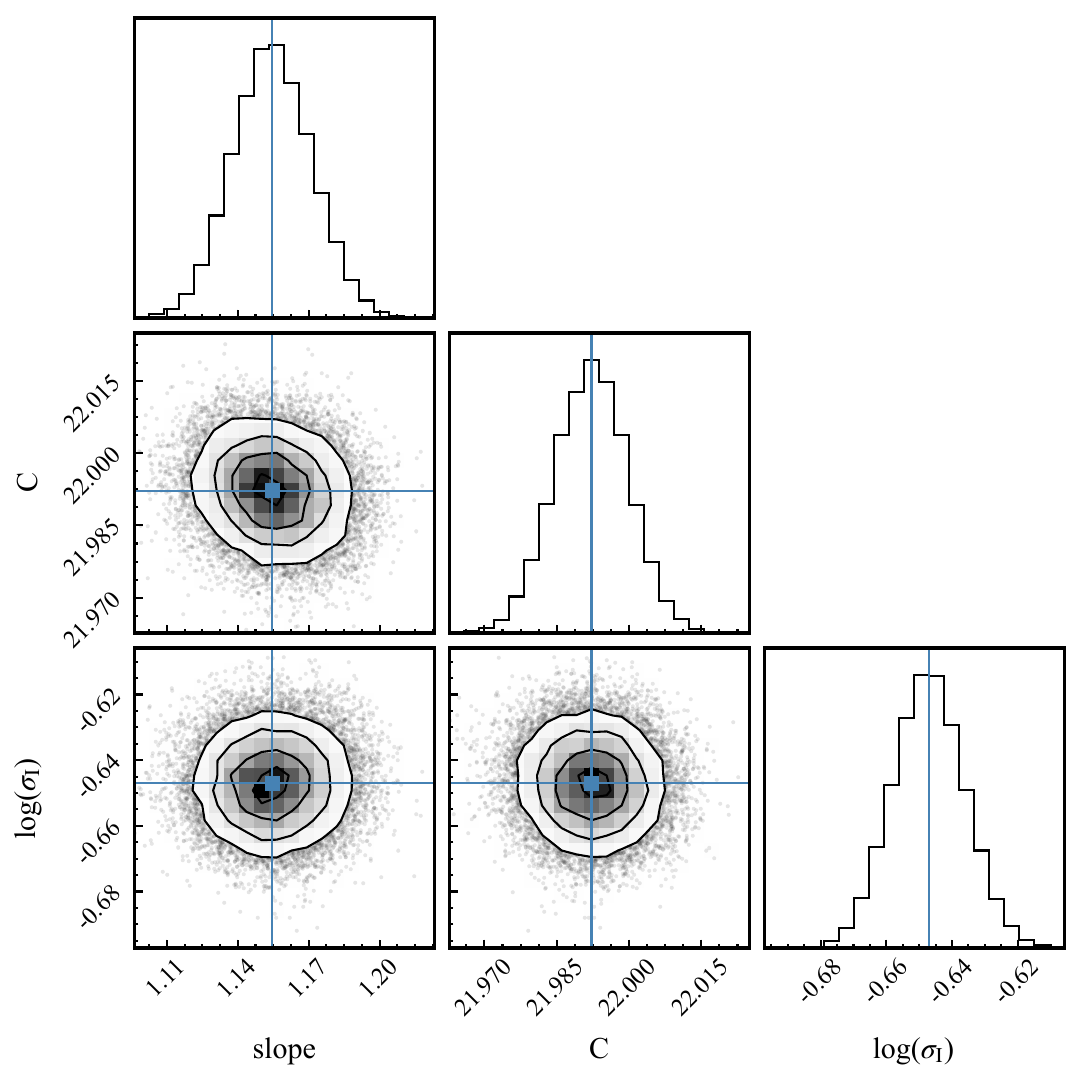}
  \caption{The Markov chain Monte Carlo sampling of our fitting for the $L_{\rm 144MHz}$-SFR relation. Three parameters are the slope, the constant value, and the intrinsic scatter in logarithm scale, respectively.}
     \label{fig:corner}
\end{figure} 

Then we visually check the LoTSS image of these galaxies and exclude those showing jet-like features and require the signal-to-noise ratio of radio detection larger than 5.
We note that this visual classification cannot exclude unresolved radio AGN. However, they are mostly quiescent galaxies in low redshift (e.g., Fig.~\ref{fig:sfms}) and thus will be excluded from the SFG sample by the H$\alpha$ equivalent width cut.
Considering that there are only very few unresolved radio AGN in the final SFG sample (e.g., some outliers in Fig.~\ref{fig:sfms}), the radio AGN contamination for the fitting should be marginal.
These unresolved radio AGN could be one source of the intrinsic scatter of the relation and could make the relation slightly overestimated.
This way we get a sample of 906 secure SFGs with reliable SFR and radio luminosity measurements.

Fig.~\ref{fig:dbpt} shows the emission line selection of this SFG sample.
$\Delta$BPT is the distance to the division line in the BPT diagram \citepads[Equation 1 in][]{2003MNRAS.346.1055K},
and the negative values represent that galaxies locate in the star-forming region.
As a result of all the constrains, this SFG sample has a SFR limit about 0.1 $\rm M_{\odot}\, yr^{-1}$ (see Fig.~\ref{fig:l144-sfr}),
thus our result about $L_{\rm 144MHz}$-SFR relation may not be applicable for SFGs with SFRs lower than this value.

We set three parameters in the linear fitting of the $L_{\rm 144MHz}$-SFR relation.
In addition to the slope and constant, we introduce an intrinsic scatter parameter,
which should be the scatter level of the $L_{\rm 144MHz}$-SFR relation after excluding the effect from measurement errors.
Our fitting follows the method in \citetads{2010arXiv1008.4686H}.
The Markov chain Monte Carlo (MCMC) sampling result is shown in Fig.~\ref{fig:corner}.
The intrinsic error is found to be small, about 0.23\,dex,
which indicates that the radio emission is a good tracer for galaxy global SFR.
The relation is listed in Equation~\ref{eq:l144-sfr}.
At the 3$\sigma$ level, our result is consistent with previous results based on different SFR tracers and radio data from different frequencies \citepads{2016MNRAS.461.1898W,2017ApJ...847..136B,2018MNRAS.475.3010G,2021A&A...648A...6S,2022A&A...664A..83H,2023MNRAS.523.1729B}.
\end{appendix}

\end{document}